\documentclass[10pt,letterpaper]{article}
\usepackage{amsmath}
\usepackage{arydshln}
\usepackage{authblk}
\usepackage[utf8]{inputenc}
\usepackage[backend=biber,style=numeric-comp,sortcites=true,minnames=6,maxbibnames=7,maxcitenames=7,giveninits=true,terseinits=true,sorting=none]{biblatex}
\usepackage[singlelinecheck=false]{caption}
\usepackage{color,soul}
\usepackage{comment}
\usepackage{fancyhdr}
\usepackage[a4paper]{geometry}
\usepackage{graphicx}
\graphicspath{{Figures/}}
\usepackage[pagewise]{lineno}
\usepackage{multirow}
\usepackage{outlines}
\usepackage{parskip}
\usepackage{subcaption}
\usepackage{textcomp}
\usepackage{todonotes}
\usepackage{ulem}
\usepackage{verbatim}
\usepackage{wasysym}

\bibliography{bibliography.bib}

\DeclareFieldFormat[book, article, thesis, inproceedings]{title}{#1}
\DeclareFieldFormat{journaltitle}{#1}
\DeclareFieldFormat{booktitle}{#1}
\DeclareFieldFormat{postnote}{\mkcomprange[{\mkpageprefix[pagination]}]{#1}}
\DeclareFieldFormat{pages}{\mkcomprange{#1}}

\DeclareNameAlias{default}{family-given}

\renewbibmacro{in:}{\ifentrytype{article} {} {\printtext{\bibstring{in} \intitlepunct}}}

\renewbibmacro*{journal+issuetitle}{%
  \usebibmacro{journal}%
  \setunit*{\addperiod\addspace}%
  \iffieldundef{series}
    {}
    {\newunit
     \printfield{series}%
     \setunit{\addspace}}%
  \setunit{\addperiod\addspace}%
  \usebibmacro{issue+date}%
  \setunit{\addcolon\space}%
  \usebibmacro{issue}%
  \setunit{\addsemicolon\space}%
  \usebibmacro{volume+number+eid}%
  \newunit}
  
\DeclareFieldFormat[article,periodical]{number}{\mkbibparens{#1}}
\renewbibmacro*{volume+number+eid}{%
  \printfield{volume}%
  \printfield{number}%
  \setunit{\addcomma\space}%
  \printfield{eid}}

\renewbibmacro*{issue+date}{%
  \iffieldundef{issue}
    {\usebibmacro{date}}
    {\printfield{issue}%
     \setunit*{\addspace}%
     \usebibmacro{date}}%
  \newunit}

\makeatletter
\renewcommand{\maketitle}{\bgroup\setlength{\parindent}{0pt}
\begin{flushleft}
  \Large{\textbf{\@title}}
  
  \vspace{2ex}
  \textit{\normalsize{\@author}}
\end{flushleft}
\egroup
}

\makeatother

\title{Characterization and \textit{ex vivo} application of flexible 2D scintillating coatings in ultra-high dose rate electron beams for FLASH radiotherapy.}
\author[a,b$\dagger$]{\underline{Verdi Vanreusel}}
\author[c,d]{Stephen Brown}
\author[d]{Shujat Ali}
\author[e]{Thomas De Kerf}
\author[d]{Anthony J. Doemer}
\author[f]{Paul Leblans}
\author[d]{Benjamin Movsas}
\author[d]{Humza Nusrat}
\author[g]{Behzad Shirmard}
\author[d]{Kundan Thind}
\author[e]{Steve Vanlanduit}
\author[b,h]{Dirk Verellen}
\author[h]{Alessia Gasparini}
\author[a]{Luana de Freitas Nascimento}
\affil[a]{Research in Dosimetric Applications, SCK CEN, Boeretang 200, 2400 Mol, Belgium}
\affil[b]{AReRO, University of Antwerp, Universiteitsplein 1, 2610 Wilrijk, Belgium}
\affil[c]{Department of Radiation Oncology, Henry Ford Health, Michigan State University, Detroit, MI, USA}
\affil[d]{Department of Medical Physics, Wayne State University, Detroit, MI, USA}
\affil[e]{InViLab, University of Antwerp, Groenenborgerlaan 171, 2020 Antwerpen, Belgium}
\affil[f]{Innovation office, Agfa N.V., Septestraat 27, 2640 Mortsel, Belgium}
\affil[g]{BrightComsol, Vienna, Austria}
\affil[h]{Department of Medical Physics, Radiotherapy, Iridium Netwerk, Oosterveldlaan 22, 2610 Wilrijk, Belgium}

\begin{document}
\maketitle
\vspace{5ex}
\section*{Abstract} 
The increasing interest in FLASH-RT has lead to both the development of new and the conversion of existing linear accelerators to enable ultra-high dose rate (UHDR) irradiations for preclinical research. 
Dosimetry in this setting remains challenging with several crucial aspects missing.
This work shows the challenges and needs for real-time 2D UHDR dosimetry, and aims to present a solution in the context of preclinical irradiations in non-homogeneous UHDR electron beams.
An experimental camera-scintillation sheet combination, was used to investigate the spatial dose distribution of a converted UHDR Varian Trilogy™ (16 MeV; 25x25 cm²). The effect of triggering was investigated and preprocessing actions developed. The dosimetric system was characterized by variation of the number of pulses and source to surface distance (SSD) and its application was investigation by variation of bolus thickness and ambient light intensity. The challenges of prelcinical real time 2D dosimetry with scintillating coatings were assessed by ex vivo irradiations of a 1) rat brain, 2) mouse hindlimb and 3) whole body mouse. The SSD, field size and number of pulses were, respectively, 80 cm, 0.6x0.6-2x2 cm² and 1-90 pulses for the rat brain; 100 cm, 10x10cm² and 23 pulses for the mouse hindlimb, and 100 cm, 20x20 cm² and 6 pulses for the whole body mouse. Radiochromic EBT XD film was used as passive reference dosimeter. 
The coating showed a linear response with the number of pulses in the absence (R²\textgreater0.998) and presence (R²\textgreater0.995) of transparent bolus, up to 3 cm thick, and with the inverse squared SSD (R²\textgreater0.963). The presence of ambient light decreases the signal-background ratio from 32.6 (dark) to 4.9 (maximum intensity). The sheet showed to have sufficient flexibility to be molded on the subjects' head/hindlimb/holder, following its curvatures. The use of a single camera with curved surfaces resulted in blind spots. Linearity with number of pulses was preserved in a preclinical setting. For small field sizes the light output became too low, resulting in noisy dose maps. The system showed robust within 5\% for camera set up differences. Calibration of the system showed to be complicated due to set up variations in combination with the inhomogeneity of the beam. 
We showed the need for 2D real-time dosimetry to determine beam characteristics in non-homogeneous UHDR beams using a preclinical FLASH radiotherapy setting. We presented one solution to meet this need with scintillating based dosimetry.

\vfill

\flushleft{$^\dagger$\textit{Corresponding author:} verdi.vanreusel@sckcen.be \newline \textit{Address:} Boeretang 200, 2400 Mol, Belgium}\\\vspace{2ex}
\textit{\textbf{Keywords:} FLASH radiotherapy; UHDR dosimetry; real time dosimetry; 2D dosimetry; Scintillating sheets; preclinical} 
\newpage

\setcounter{page}{1}
\section{Introduction}
A growing interest in the FLASH effect, the biological observation that damage to irradiated tissue is reduced  when delivering the dose at ultra-high dose rate (UHDR), has led to an increased availability of UHDR irradiation sources. While the initial research was limited to dedicated experimental electron linacs~\cite{favaudon2014ultrahigh}, UHDR modalities are now available on synchrotrons, cyclotrons, linacs for intra-operative electron radiotherapy and converted linacs for external beam radiotherapy~\cite{Farr2022}. The varied modalities have different beam structures characterized by their beam parameters. 
In general, beams are quasi-continuous or pulsed, referring to their duty cycle, and are delivered broad beam or scanned beam. Two characteristic beam parameters in quasi-continuous UHDR beams are the average dose rate (aDR) and total delivery time. In pulsed beams, also the dose per pulse (DPP) and dose rate within the pulse (iDR) are relevant~\cite{Farr2022, bohlen2024recording}. For scanned beams the definition of average dose rate is not straight forward and a topic of research on its own as multiple spots contribute to the dose at a point~\cite{deffet2023definition}. By general consensus, a beam is considered UHDR when the average dose rate exceeds $\sim$ 40 Gy/s, whereas a beam is referred to as "FLASH" if the FLASH effect is observed. The FLASH effect might be tissue dependent, meaning that a beam can theoretically be FLASH for one tissue type and not for another. Also, there is growing evidence that there is a dose threshold under which the FLASH effect disappears~\cite{vozenin2020all}. Since the beam parameter requirements to obtain the FLASH effect remain a topic of research, they need to be well characterized and reported~\cite{bohlen2024recording}.
\newline
Many preclinical experiments have validated the FLASH effect for various tumor and tissue types, using different beam modalities~\cite{bourhis2019clinical, diffenderfer2020design, sorensen2022vivo, vozenin2022towards}. However, most of the studies ignore time considerations relying on passive or integrated point dosimetry. This information is crucial to determine the mechanisms underlying the FLASH effect, an understanding of which is important in optimizing a differential cancer to normal tissue response. In addition, for non-homogeneous beams, the dose and dose rate distributions are essential quantities to characterize. In such beams, the use of point dosimeters is more complex and positioning of dosimeters becomes extremely important. Therefore, real-time 2 dimensional dosimetry is needed, particularly for the proper interpretation of biological experiments. This is especially relevant for converted linacs, as these usually have a Gaussian beam profile due to removal of the flattening filter to obtain UHDR.  
\newline
Research groups are investigating arrays of dosimeter elements~\cite{Yang2022, Clark2024, Schonfeld2024A, vasyltsiv2024design}, radiation-induced acoustic imaging~\cite{Bjegovic2024}, Cherenkov dosimetry~\cite{rahman2021spatial}, and scintillating sheets with a camera~\cite{rahman2020characterization, ashraf2021single, rahman2021spatial, Clark2023, goddu2024high, Kanouta2024, Levin2024, Rieker2024, vanreusel2024dose}. Arrays have the additional challenge of being held in a matrix and having gaps between the elements, a limiting factor for beams with strong gradients. Cherenkov dosimetry and scintillating sheets have been extensively investigated in beams with conventional dose rates for in vivo dosimetry and patient specific quality assurance~\cite{Nascimento2025ReviewRealTime2D, darafsheh2024radioluminescence}. While not yet commercially available, promising results have been obtained, with average gamma passing rates of 97.4\% (3\%/3 mm) for clinical plans with the treatment planning system as ground truth~\cite{ando2021verification}. In the context of UHDR, 2D scintillating dosimetry has been mainly investigated for pencil beam scanning proton beams~\cite{rahman2020characterization,Clark2023,goddu2024high,Kanouta2024,Levin2024,Rieker2024}. 
\newline
In this study, two flexible Yttrium Aluminium Garnet (YAG) scintillating sheets, in combination with two CMOS cameras, are characterized and investigated in a preclinical setting for a non-homogeneous UHDR clinical linac electron beam converted to allow for UHDR.  
One of the benefits of YAG is its short decay time, which makes it especially useful for pulsed UHDR dosimetry. A point scintillator~\cite{Vanreusel2024In-vivo} and scintillating sheets based on YAG have already been characterized in UHDR electron beams~\cite{vanreusel2024dose} and  the latter has also been tested in very high energy electron (VHEE) beams~\cite{Rieker2024}.
\section{Materials and methods}
\subsection{Dosimetry system}
The dosimetry system consists of a camera and scintillating sheet. The scintillating sheet emits light, proportional to the radiation dose rate and the camera collects and digitizes the light to a two-dimensional grid of grey values. In the current study scintillating sheets consisting of YAG (Y$_{3}$Al$_{5}$O$_{12}$:Ce$^{3+}$) crystals embedded in a silicon matrix, with a particle loading of 30\% were characterized. Two sheets were tested. The "sample 1" and "sample 2" sheets have a size of 5x5 cm² and varied in silicon composure, thickness (200 and 190 µm, respectively) and flexibility. For comparison, the YAG sheet used in Vanreusel et al.~\cite{vanreusel2024dose} was also exposed during the experiments ("sample 3").
\newline
Two cameras were evaluated. The first camera, a C-blue One (Oxford Imaging, USA), is a scientific CoaXPress 2.0 camera with an IMX426 (Sony, Japan) CMOS sensor of size 816x624 pixels², cooled to -30 °C. The second camera, a acA640-750um (Basler) (Basler, Netherlands), is a compact USB 3.0 camera with PYTHON 300 (onsemi, USA) CMOS sensor with a size of 640x480 pixels². These cameras allow high temporal resolution recordings with maximal frame rates of 1594.7 and 751.9 Hz, respectively. Both cameras use a global shutter and allow free-running and triggered frame recording. In free-running recording, frames are acquired continuously, whereas triggered recording requires an external trigger signal to acquire a frame.

\subsubsection{Preprocessing and analysis}
\label{sec: preprocessing and analysis}
The obtained images were preprocessed and analyzed using Python scripts in Fiji (ImageJ) and JupyterLab according to the following steps:
\begin{enumerate}
    \item projective transformation,
    \item noise removal,
    \item signal extraction,
    \item background subtraction, and
    \item ramp up correction.
\end{enumerate}

The projective transformation was used to correct for the non-zero angle between the straight line from the camera to the target and the straight line between the radiation source to the scintillating sheets. The perspective transformation matrix was obtained by warping 4 corners of a flat grid pattern (i.e. graph paper) to a rectangular grid.  The perspective transformation was then applied to all recorded images, resulting in a beam's eye view of the irradiation field from the camera’s perspective. 
\newline
Noise removal consisting of median filtering with a radius of 2 pixels and a threshold of 50 was performed. 
\newline 
For the signal extraction, the signal of the scintillating sheet was defined as the average grey value of a region of interest (ROI) in the center sheet.
\newline
Background subtraction was a two-step process in which the background was first extracted and then subtracted from the signal. The background subtraction method differed for triggered and for free running acquisitions. In the former, background subtraction was performed per frame, with the background defined as the average grey value of a ROI outside the scintillating sheet. In the latter, the signal ROI was used to construct a time-trace consisting of the average grey value per frame. The background was defined as the average of all data points from the time trace prior to the first pulse. This background was subtracted from every point of the time-trace. The process can be described as follows.
$$
    S'(x)= 
\begin{cases}
    S(x) - BG(x),& \text{for triggered acquisition}\\
    S(x) - BG,   & \text{for free running acquisition.}
\end{cases}
$$
where, $S'(x)$ is the background subtracted signal for frame $x$, $S(x)$ is the raw signal for frame $x$, $BG(x)$ is the average grey value of a ROI outside the scintillating sheet for frame $x$, and $BG$ is the mean of the average grey value from the signal ROI for all frames prior to the first pulse.
\newline
As the triggered acquisition showed an initial ramp up of the signal-per-pulse, caused by overlapping signal due to the decay time of the scintillating sheet (see sections \ref{sec: Results setup optimization} and \ref{sec: Discussion setup optimization}), an additional (ramp up) correction was introduced. This correction can be described by:
$$S''(x) = S'(x) \cdot k(x).$$
where $S''(x)$, $S'(x)$, and $k(x)$ are the corrected signal, background subtracted signal and correction factor, for frame $x$, respectively. 
To obtain $k(x)$, a representative normalized time trace was constructed. This was an average of 6 time traces where 40-90 pulses were delivered, normalized by the respective signals of the last pulse. Then $k(x)$ was defined as the inverse of this time trace as successively explained in Figure \ref{fig: RampUpCorrection_All}. The correction factor $k(x)$ was dependent on the camera settings (e.g. iris opening, gain, etc.).

\subsection{Characterization}
\subsubsection{Irradiation setup}
\begin{figure}[!ht]
    \centering
    \includegraphics[width=.8\linewidth]{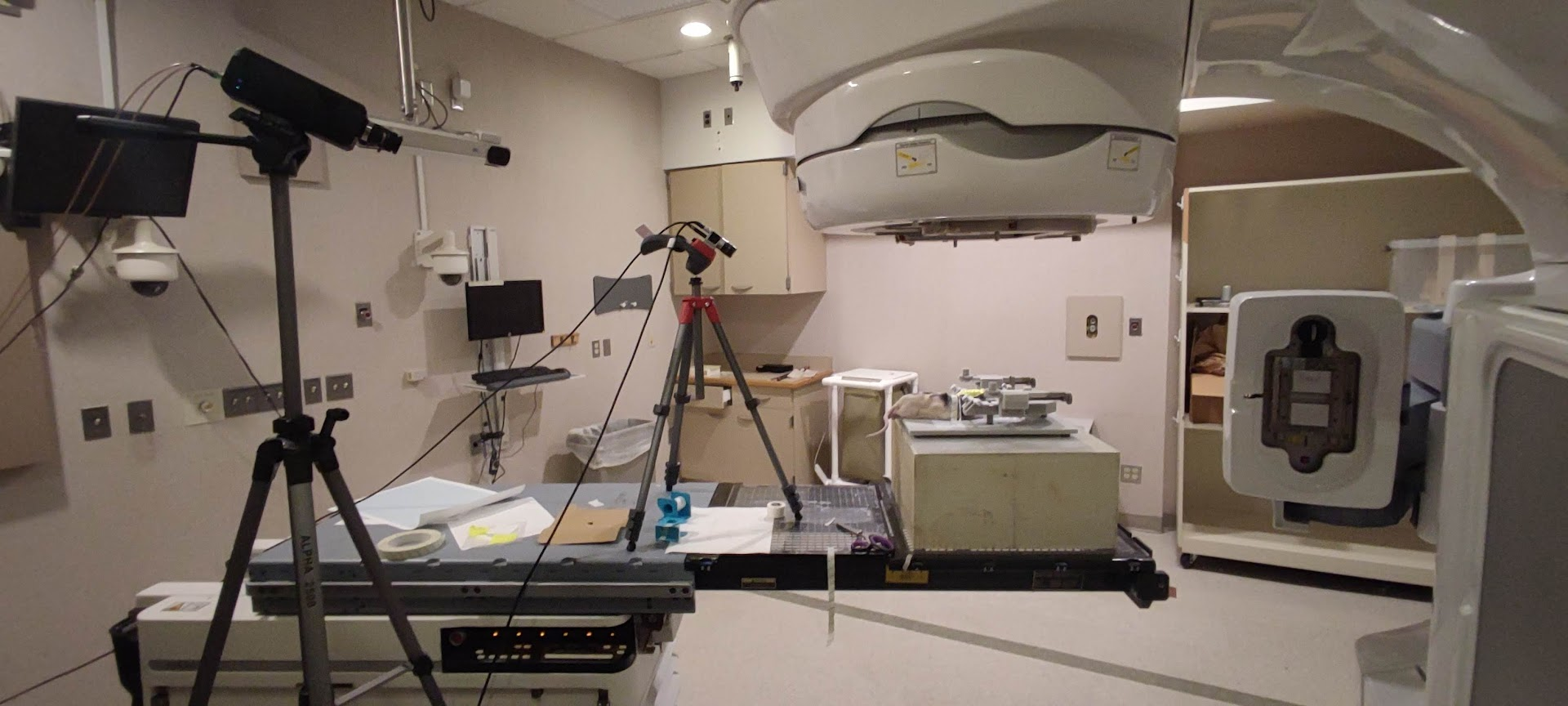}
    \caption{The camera setup with both cameras positioned as high as possible to minimize the angle between the camera-isocenter axis and the central axis of the beam. The C-blue and Basler cameras were positioned next to and on top of the treatment couch, respectively.\label{Fig: Camerasetup}}
\end{figure}

A converted Varian Trilogy™ (Varian, USA) was used to deliver a 16 MeV UHDR electron beam with a Gaussian beam profile, an average pulse repetition frequency of 115 Hz and dose per pulse of 0.78 Gy in the center of the field for a source to surface distance (SSD) of 100 cm (dose at the surface with no build-up). The sheets were positioned side by side, together with a piece of radiochromic film (Gafchromic\texttrademark\ EBT3, Ashland Advanced Materials, Bridgewater, NJ), similar in size, in the center of a 25x25 cm² field on top of a rectangular solid water phantom. The C-blue and Basler cameras were, respectively, positioned next to- and on the couch, as high as possible to minimize the angle between the camera-isocenter axis and the central axis of the beam. Both free-running and triggered recording were investigated. During the former, the C-blue and Basler cameras operated at a frame rate of 554.9 and 1000.0 kHz, respectively, with an exposure time of 1 ms. During triggered frame recording the exposure time was 2.5 ms. A picture of the camera setup is shown in Figure~\ref{Fig: Camerasetup}.

\subsubsection{Irradiations}
Irradiations under varying conditions were performed to characterize the:
\begin{itemize}
    \item dose linearity,
    \item effect of bolus,
    \item effect of aDR and DPP variation, and
    \item influence of ambient light in the room on the system.
\end{itemize}
At the time of dose linearity assessment, the use of triggering also was investigated. Based on the advantages of triggering to reliably measure signal, the decision was made to use triggered acquisition for subsequent irradiations. The irradiation parameters are summarized in Table~\ref{tab: irr params}. The bolus material consisted of 1.15 cm thick transparent polymer gel slabs (Clearsight Bolus, Radiation Products Desing, Albertville, MN; note that 1.15 cm thick bolus is equivalent to 1.0 cm water). During all tests, the light was off except for the ambient light experiment, where the light dimmer was divided in five equidistant steps between off and maximal intensity. The light switch positions were labeled 0-4 with 0 corresponding to lights off and 4 corresponding to lights at maximal intensity.
\newline
Since the position of the film between irradiations was subject to small variations (\textless1 cm), in combination with the non homogeneity of the beam, the number of pulses were used as surrogate measure for total dose. The radiochromic film was used to evaluate discrepancies in dose delivery.

\begin{table}[!ht]
    \centering
    \caption{The irradiation parameters used for the characterization of the sheets.}
    \resizebox{\textwidth}{!}{
    \begin{tabular}{cccccc}
         Investigation of& \# Pulses & SSD & Bolus & Field size & Position dimmer  \\
         && [cm] & [cm] & [cm²] & \\
         \hline
         Dose linearity& 1 - 90 & 100  & 0  & 25x25  & off \\
         Effect of Bolus & 1 - 90 & 97 - 99  & 1 - 3  & 25x25  & off \\
         Effect of aDR and DPP & 5, 10, 15, 20 & 80 - 125  & 0  & 25x25  & off \\
         Influence of ambient light & 23 & 100  & 0  & 25x25  & off - 4 \\
    \end{tabular}}
    \label{tab: irr params}
\end{table}

\subsection{Use in preclinical environment}
The characterized sheets and cameras were tested in three ex vivo situations, relevant for preclinical research. These included rat brain~\cite{brown2015mri,jenrow2013selective,jenrow2011combined,jenrow2010ramipril,kim2004modification}, mice hindlimb~\cite{brown2019novel,kim2012plerixafor,yan2008mitigation,kumar2008radiation,kohl2007differential} and whole-body mouse~\cite{valeriote2024novel,brown2010antioxidant,brown2008histone} irradiations. In the former two, the "sample 2" sheet was used and molded directly on the skin. In the latter, the "sample 1" sheet was used and molded on a dedicated mouse holder (modified 50 mL centrifuge tube with air holes). The performed irradiations and parameters are listed in Table~\ref{tab: preclin irr params}.

\begin{table}[!ht]
    \centering
    \caption{The irradiation parameters used for the ex vivo tests.}
    \resizebox{\textwidth}{!}{
    \begin{tabular}{ccccccc}
         Investigation of& \# Pulses & SSD & Bolus & Field size & Preclinical situation & Sheet \\
         && [cm] & [cm] & [cm²] & &\\ 
         \hline
         Dose linearity& 1 - 90 & 80 & 0  & 2x2  & Rat brain & "sample 2" \\
         Effect of bolus & 30 & 80 & 1 - 3  & 2x2  & Rat brain & "sample 2" \\
         Field size& 30 & 80 & 0  & 0.6x0.6 - 2x2  & Rat brain & "sample 2" \\
         Effect of positioning & 23 & 80 & 0  & 2x2  & Rat brain & "sample 2" \\
         repeatability & 23 & 100 & 0  & 10x10  & Mice hindlimb & "sample 2" \\
         Whole body mouse & 6 & 100 & 0  & 20x20  & Whole body mouse & "sample 1" \\
    \end{tabular}}
    \label{tab: preclin irr params}
\end{table}

\section{Results}
\subsection{Setup optimization}
\label{sec: Results setup optimization}
\begin{figure}[!ht]
    \begin{subfigure}{.5\textwidth}
        \centering
        \includegraphics[width=\linewidth]{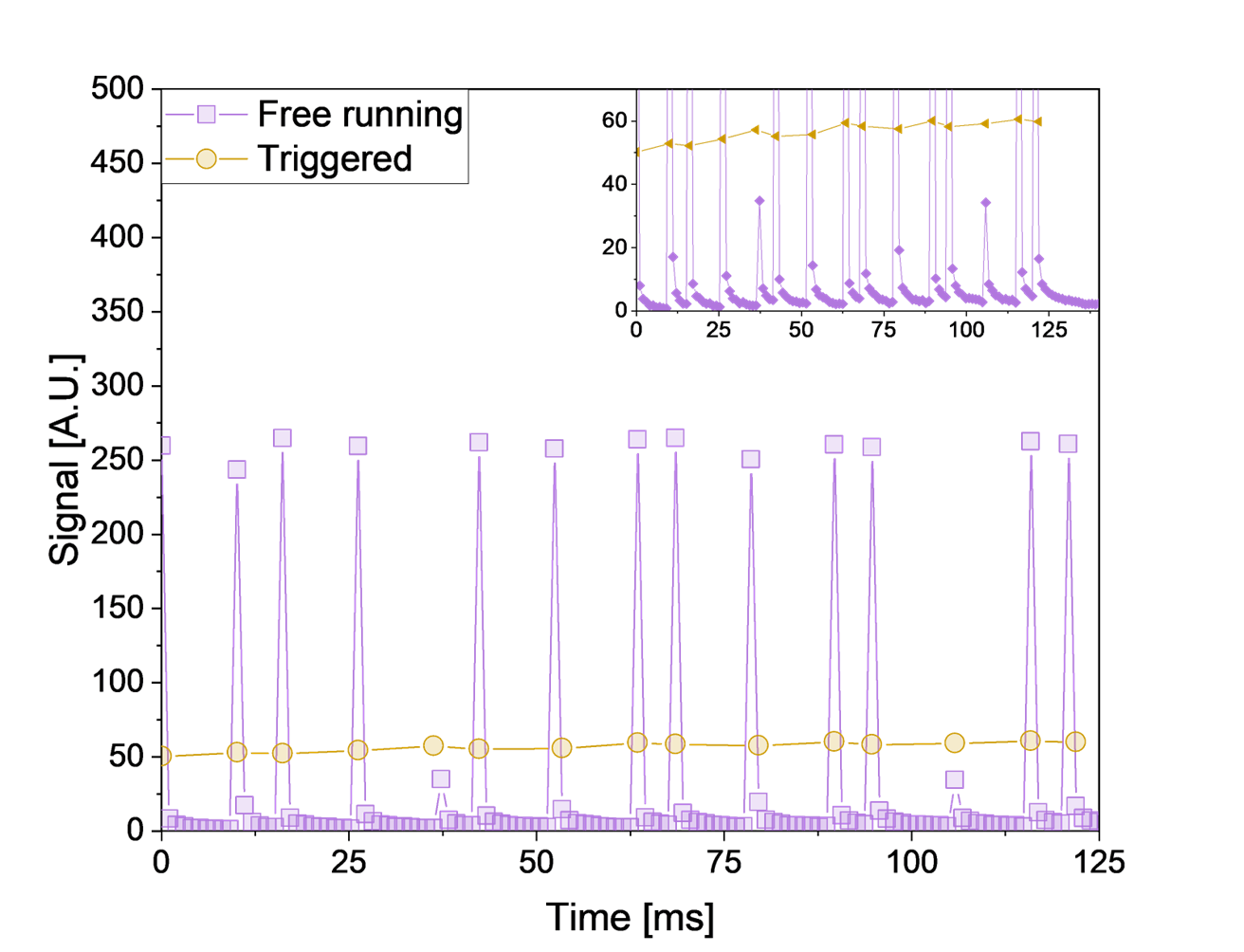}
        \caption{\centering\label{fig: TriggervsNonTrigger}}
    \end{subfigure}
    \begin{subfigure}{.5\textwidth}
        \centering
        \includegraphics[width=\linewidth]{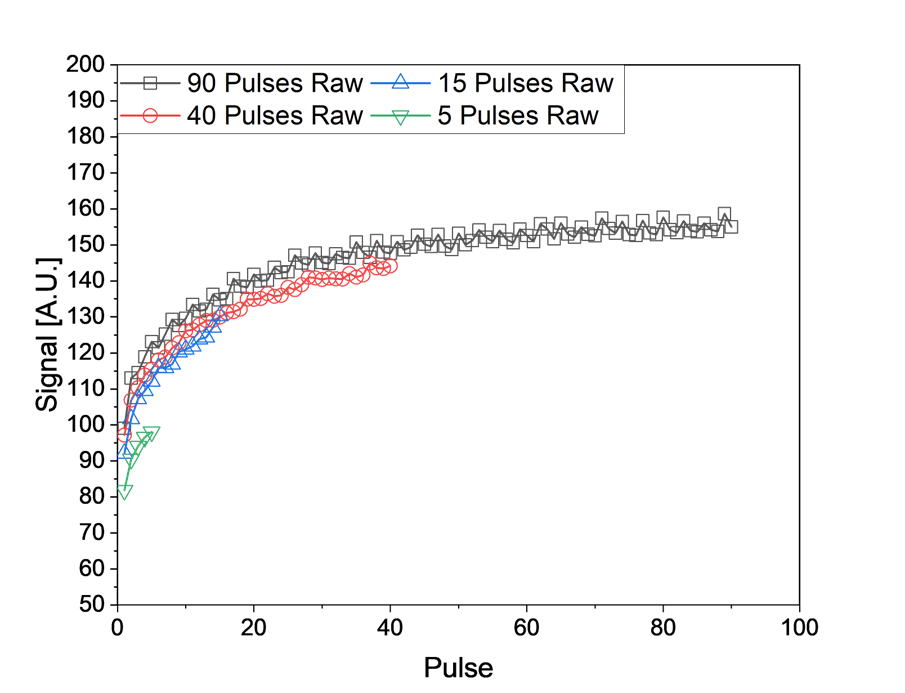}
        \caption{\centering\label{fig: RampUp}}
    \end{subfigure}
    \caption{(a) Time traces of a free running- (purple squares) and triggered- (beige circles) recording for an irradiation of 15 pulses. The location of the pulses in triggered mode were time-matched with the ones in free running mode for comparison. A zoomed in view is provided by the insert, in which the decay can be observed. (b) Time traces of triggered recordings for different amount of delivered pulses.}
\end{figure}

\begin{figure}[!ht]
    \begin{subfigure}{.5\textwidth}
        \centering
        \includegraphics[width=\linewidth]{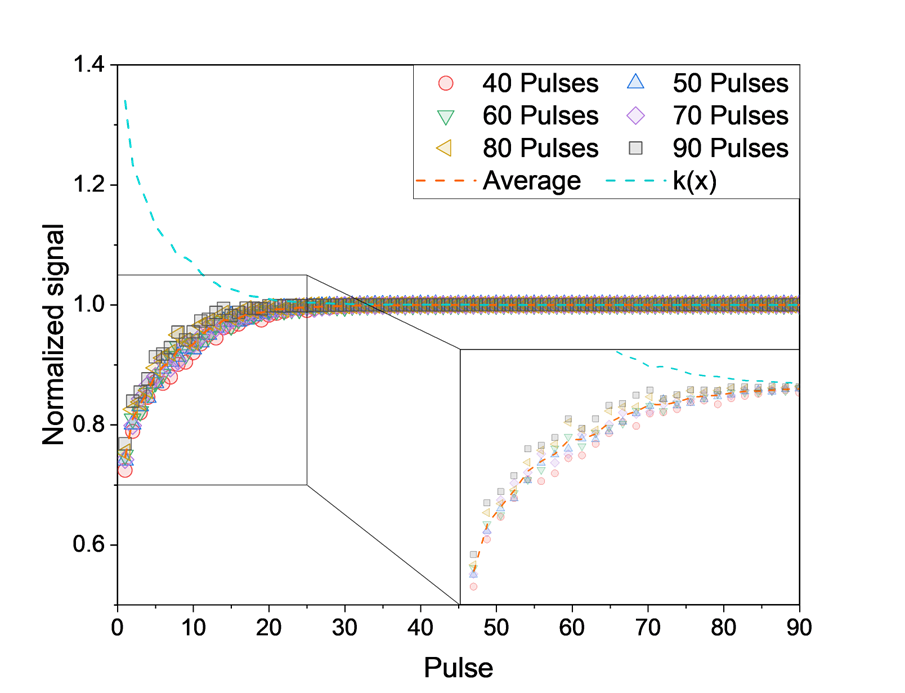}
        \caption{\centering\label{fig: RampUpCharacterizationAndCorrection}}
    \end{subfigure}
    \begin{subfigure}{.5\textwidth}
        \centering
        \includegraphics[width=\linewidth]{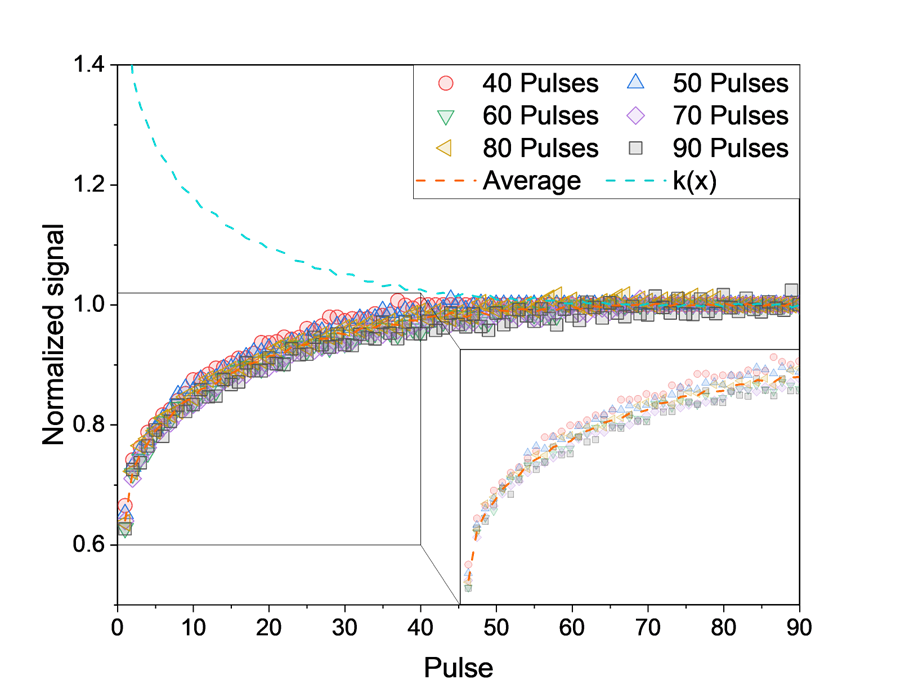}
        \caption{\centering\label{fig: RampUpCharacterizationAndCorrection_Basler}}
    \end{subfigure}
    \begin{subfigure}{.5\textwidth}
        \centering
        \includegraphics[width=\linewidth]{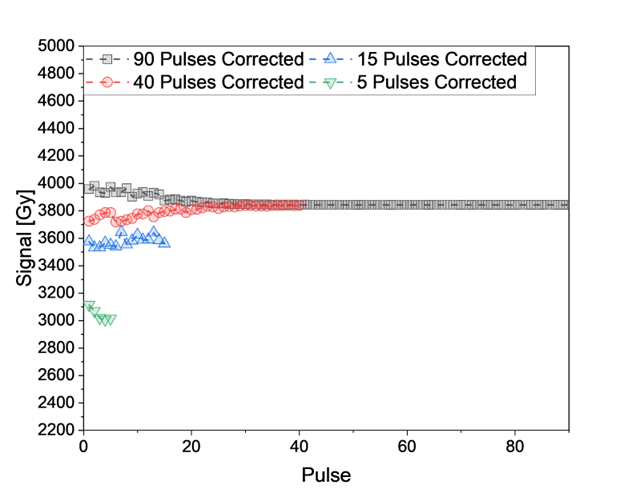}
        \caption{\centering\label{fig: RampUpAfterCorrection}}
    \end{subfigure}
    \begin{subfigure}{.5\textwidth}
        \centering
        \includegraphics[width=\linewidth]{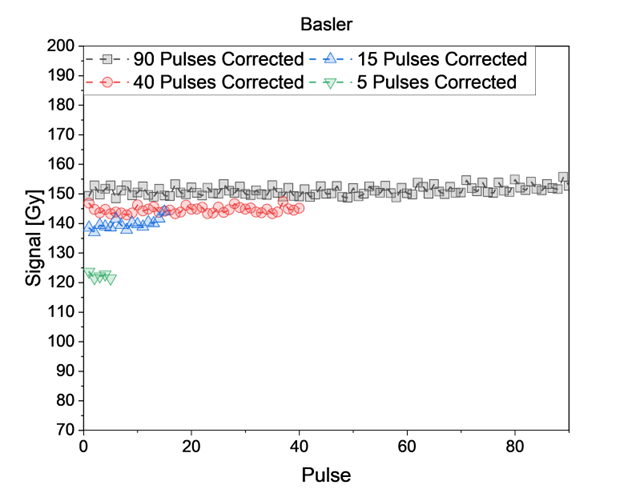}
        \caption{\centering\label{fig: RampUpAfterCorrection_Basler}}
    \end{subfigure}
    \caption{ The upper graphs show the normalized time traces for a delivery of 40, 50, 60, 70, 80 and 90 pulses, the representative normalized time trace (Average), and its inverse, which serves as ramp up correction function (k(x)) for the "sample 1" sheet in combination with the (a) C-blue and (b) Basler cameras. The lower graphs show the time traces for a delivery of 5, 15, 40 and 90 pulses after correction for the "sample 1" sheet in combination with the (c) C-blue and (d) Basler cameras.\label{fig: RampUpCorrection_All}}
\end{figure}

The first setup optimization step investigated the use of free running and triggered acquisition. Representative time traces of free running- and triggered-recordings are shown in Figure~\ref{fig: TriggervsNonTrigger}, where the triggered pulses were time-matched with the free running ones for comparison. The pulse signal was higher for most pulses in the free running setting. However, 2/15 pulses were not (fully) detected due to the non-negligible dead time of the camera. Both the background of the free running recording and the signal for the triggered setting increase in time (insert Figure~\ref{fig: TriggervsNonTrigger}). Finally, the spread on the pulse amplitude is larger for the free running recording. It was opted to use triggered acquisition for the characterization and use in the preclinical environment because of the missing pulses when using free running acquisition. Further discussion on the rationale can be found in section~\ref{sec: Discussion setup optimization}.
\newline
The signal rise was further investigated as observed for a different number of pulses in Figure~\ref{fig: RampUp}. The signal ramps up during the first ± 40 pulses after which it stabilizes. A correction was applied based on a representative normalized time trace. The representative normalized time trace, the time traces used to construct it, and its inverse, being $k(x)$ as described in section~\ref{sec: preprocessing and analysis} are shown in Figures~\ref{fig: RampUpCharacterizationAndCorrection} and ~\ref{fig: RampUpCharacterizationAndCorrection_Basler}, for the C-blue and Basler camera, respectively. The effect of the correction is shown in Figures~\ref{fig: RampUpAfterCorrection} and~\ref{fig: RampUpAfterCorrection_Basler}, for the respective cameras. The correction for the C-blue camera shows that the signal stabilizes after 20 pulses. However, closer inspection shows that this was due to saturation of the signal in most of the pixels of the ROI. Assuming that the dose per pulse was stable, this results in a suboptimal correction, where the signal of the first 20 pulses was over corrected for 5 and 90 pulses and under corrected for 40 pulses. The correction for the Basler camera was shown to be adequate with minor signal variation between different pulses. However, while no difference of signal per pulse was expected between the irradiations, an increase was observed with the delivered number of pulses for both cameras. 

\subsection{Characterization}
\begin{figure}[!ht]
    \begin{subfigure}{.5\textwidth}
        \centering
        \includegraphics[width=\linewidth]{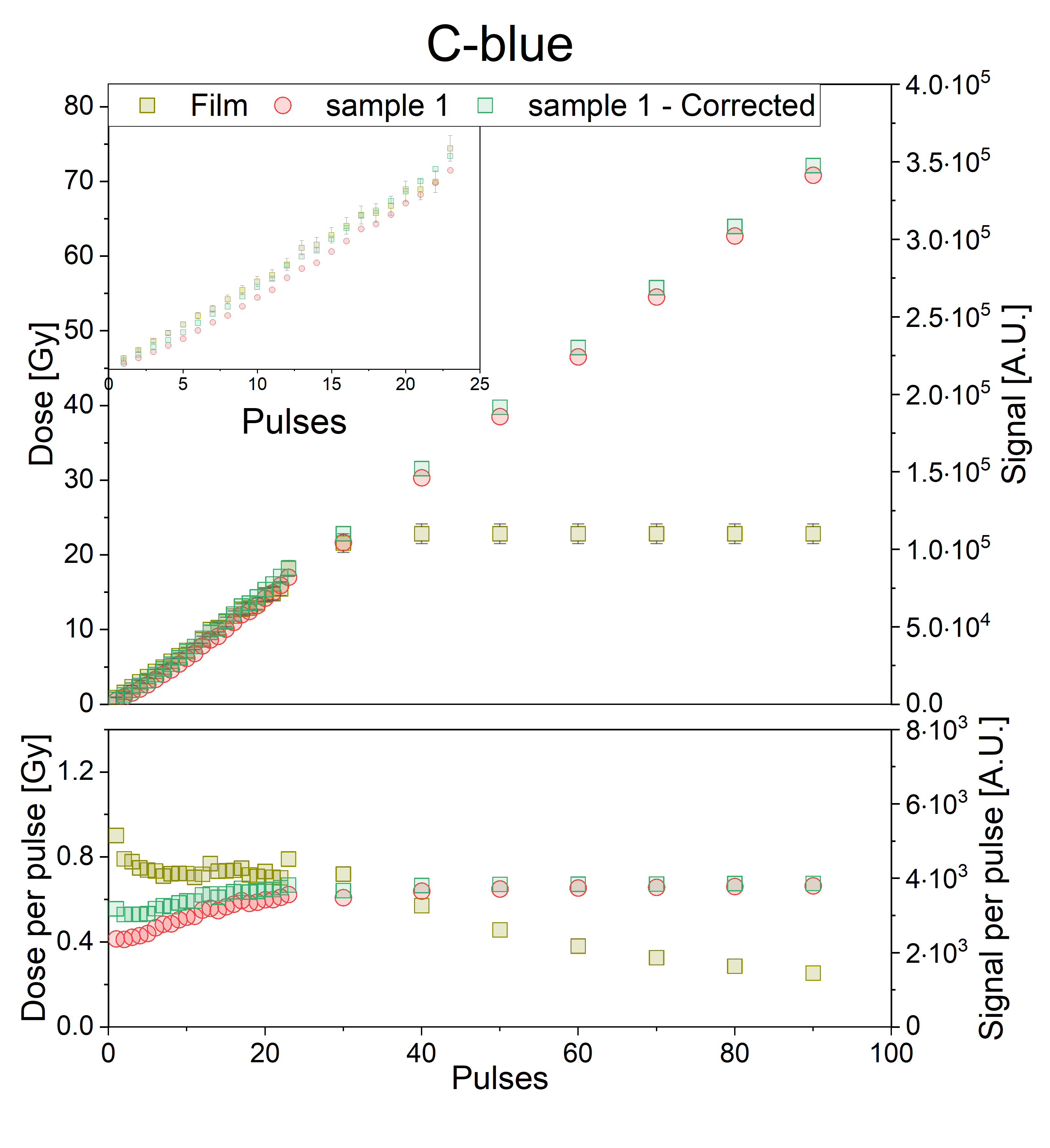}
        \caption{\centering\label{fig: PulseLinearity_Cblue}}
    \end{subfigure}
    \begin{subfigure}{.5\textwidth}
        \centering
        \includegraphics[width=\linewidth]{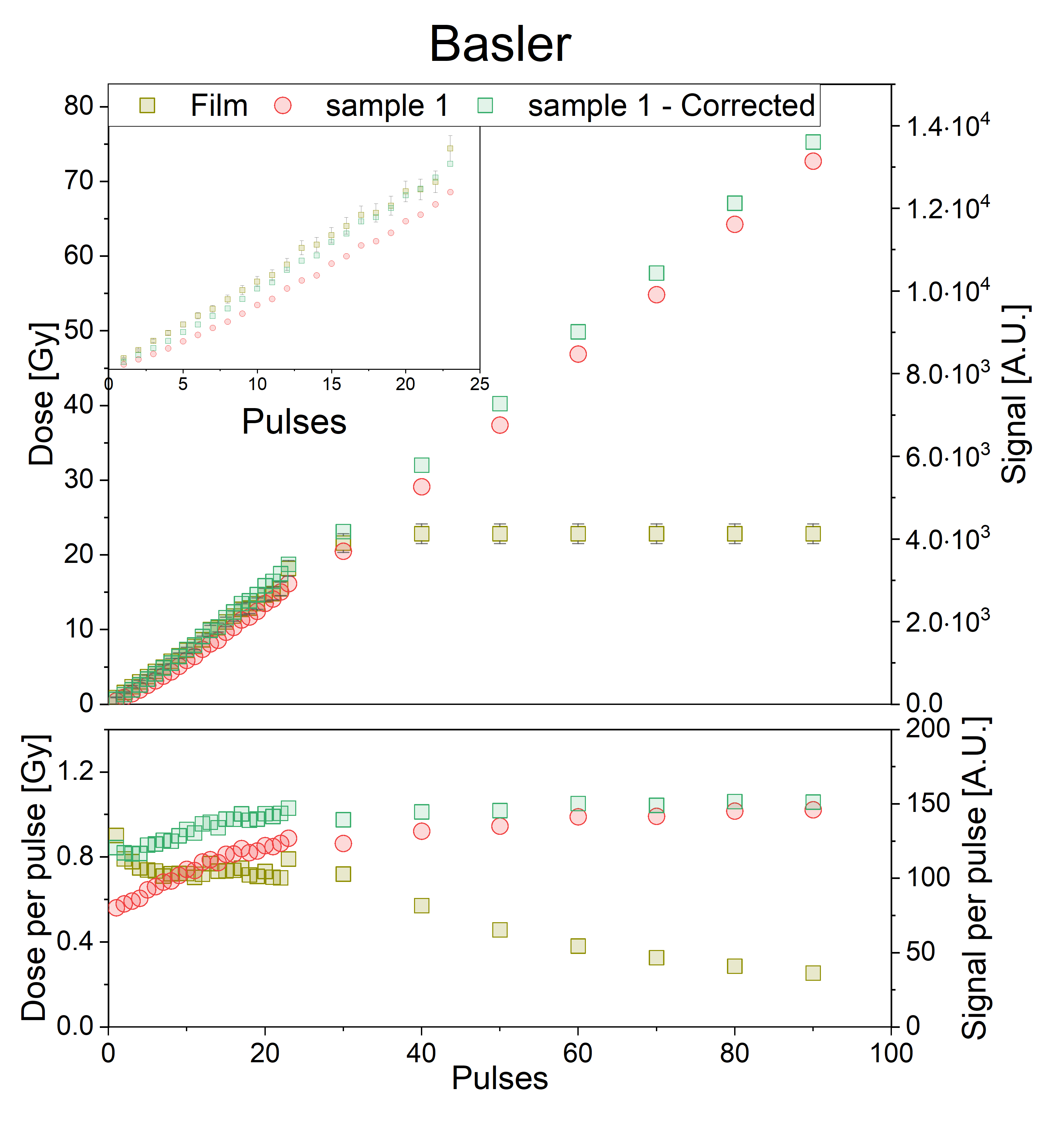}
        \caption{\centering\label{fig: PulseLinearity_Basler}}
    \end{subfigure}
    \caption{The linearity of the dosimetric system with the number of pulses, hence dose, with and without ramp up correction for the (a) C-blue and (b) Basler cameras.\label{Fig: PulseLinearity}}
\end{figure}

The dose linearity of the dosimetric system was investigated by variation of the number of pulses. The results are shown in Figure~\ref{Fig: PulseLinearity}. For clarity, only the results of "sample 1" are shown. Similar results are obtained for "sample 2" and "sample 3". Radiochromic film saturates above 21.6 Gy (30 pulses), whereas "sample 1" is linear up to at least 90 pulses (R²\textgreater0.999 for both cameras). The ramp up correction is most effective for irradiations with less pulses than needed to stabilize the signal (inserts). The lower plot in Figure~\ref{Fig: PulseLinearity} shows that despite this correction, the average signal per pulse was lower for low number of pulses, with a maximal difference of 19.5 and 22.5\% for the C-blue and Basler cameras, respectively. The opposite behavior was observed in radiochromic film where the average dose per pulse was higher for low number of pulses with a maximal difference 22.5\% in this region.

\begin{figure}[!ht]
    \begin{subfigure}{.5\textwidth}
        \centering
        \includegraphics[width=\linewidth]{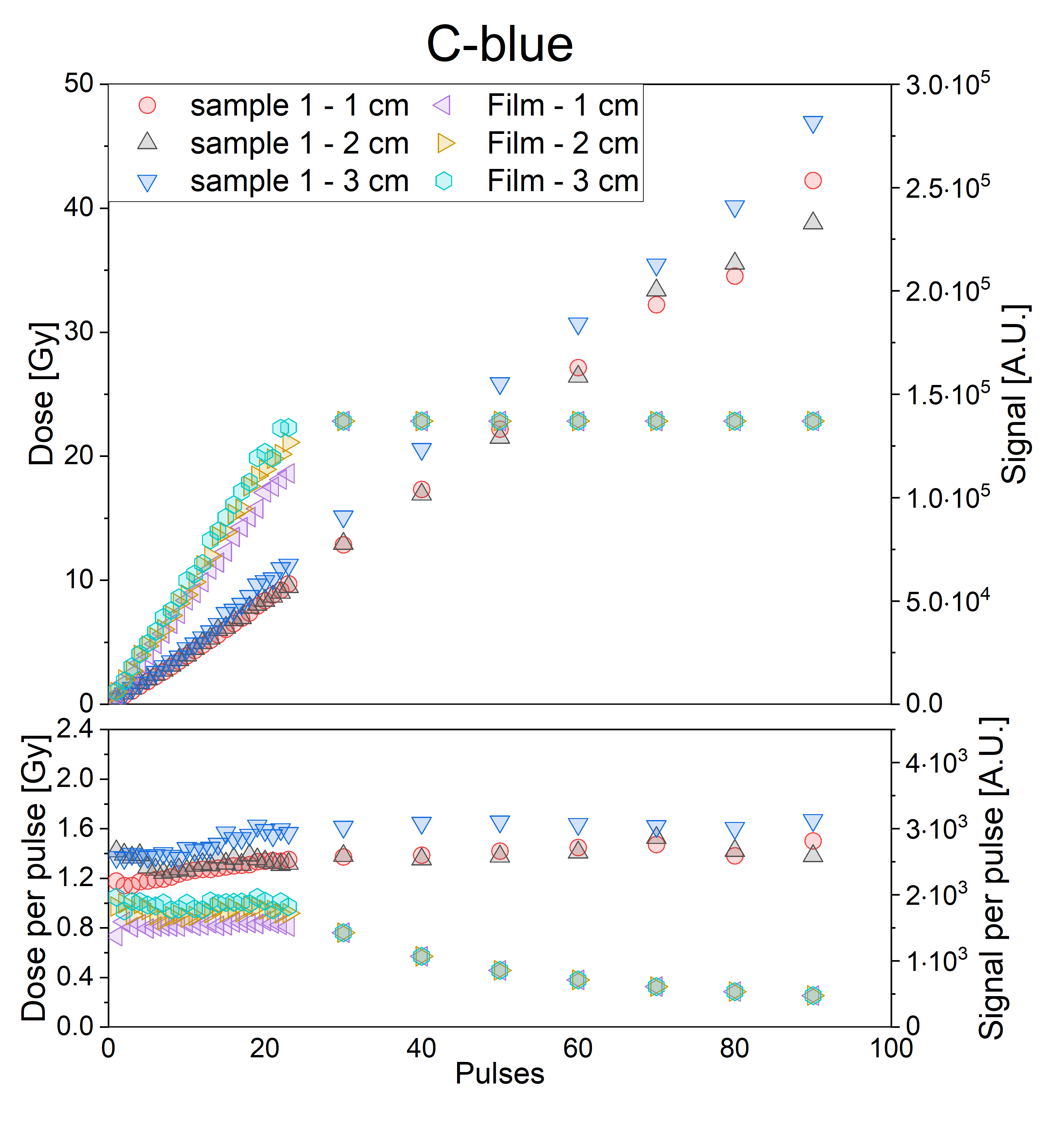}
        \caption{\centering\label{fig: Bolus_Cblue}}
    \end{subfigure}
    \begin{subfigure}{.5\textwidth}
        \centering
        \includegraphics[width=\linewidth]{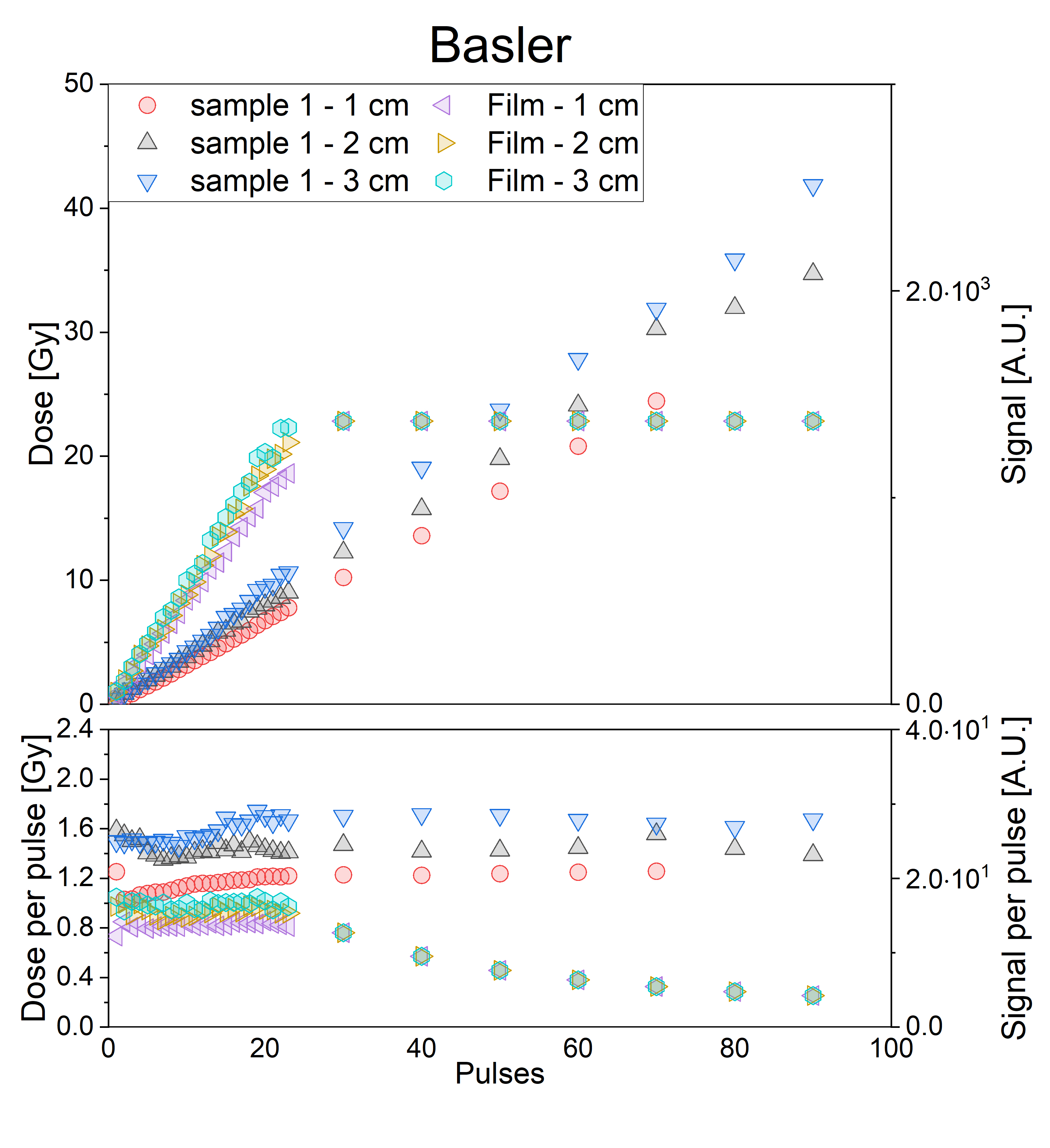}
        \caption{\centering\label{fig: Bolus_Basler}}
    \end{subfigure}
    \caption{The linearity of the dosimetric system with the number of pulses, hence dose, with different bolus thickness on top of the scintillating sheet for the (a) C-blue and (b) Basler cameras.\label{Fig: Bolus}}
\end{figure}

The effect on the dose linearity was investigated when transparent bolus slabs were added to mimic a clinically relevant situation. Figure~\ref{Fig: Bolus} shows that the dosimetric system remains linear with the number of pulses when boluses are added on top of the "sample 1" sheet (R²\textgreater0.995). Additionally, the expected dose increase with the amount of bolus due to the build up region in electron beams, was observed in both the radiochromic film and the sheet-camera system.

\begin{figure}[!ht]
    \begin{subfigure}{.5\textwidth}
        \centering
        \includegraphics[width=\linewidth]{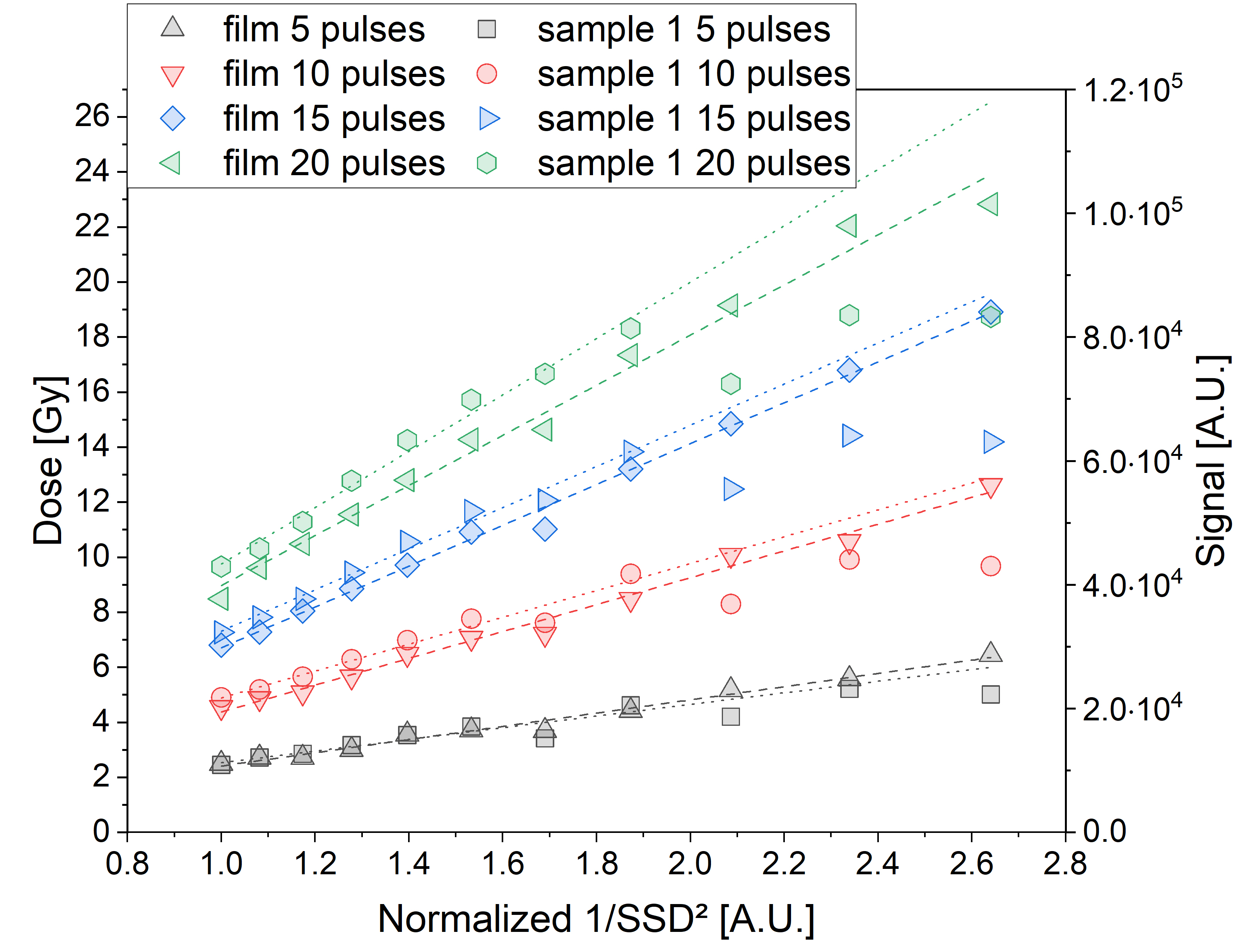}
        \caption{\centering\label{fig: SSD_Cblue}}
    \end{subfigure}
    \begin{subfigure}{.5\textwidth}
        \centering
        \includegraphics[width=\linewidth]{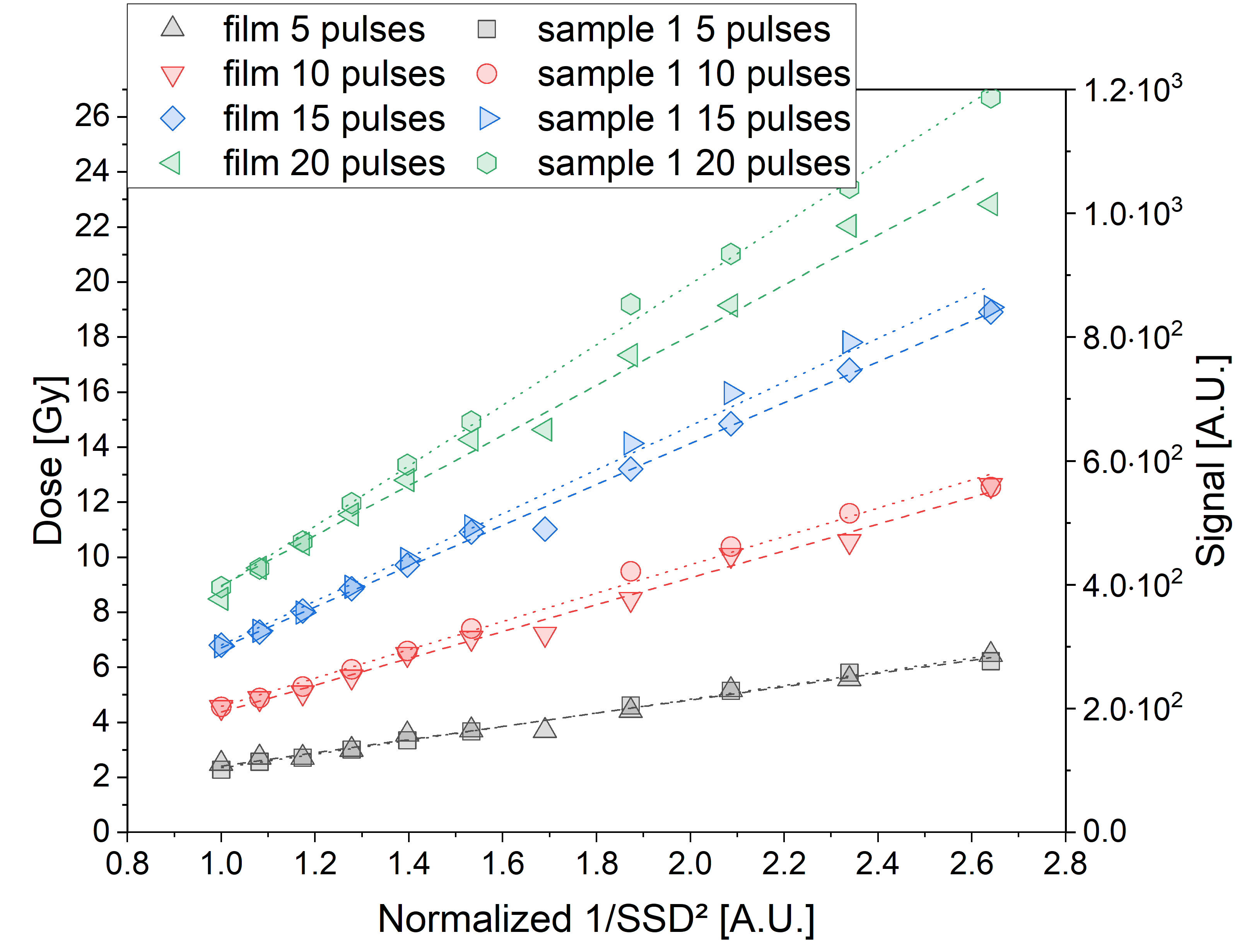}
        \caption{\centering\label{fig: SSD_Basler}}
    \end{subfigure}
    \caption{Verification of the inverse square law for the dosimetric system and radiochromic film for the (a) C-blue and (b) Basler cameras.\label{Fig: SSD}}
\end{figure}

The independence of the aDR, DPP and instantaneous dose rate (iDR) was investigated by verifying the inverse squared relationship between signal and SSD. Figure~\ref{Fig: SSD} illustrates the results.

\begin{figure}[!ht]
    \begin{subfigure}{.5\textwidth}
        \centering
        \includegraphics[width=\linewidth]{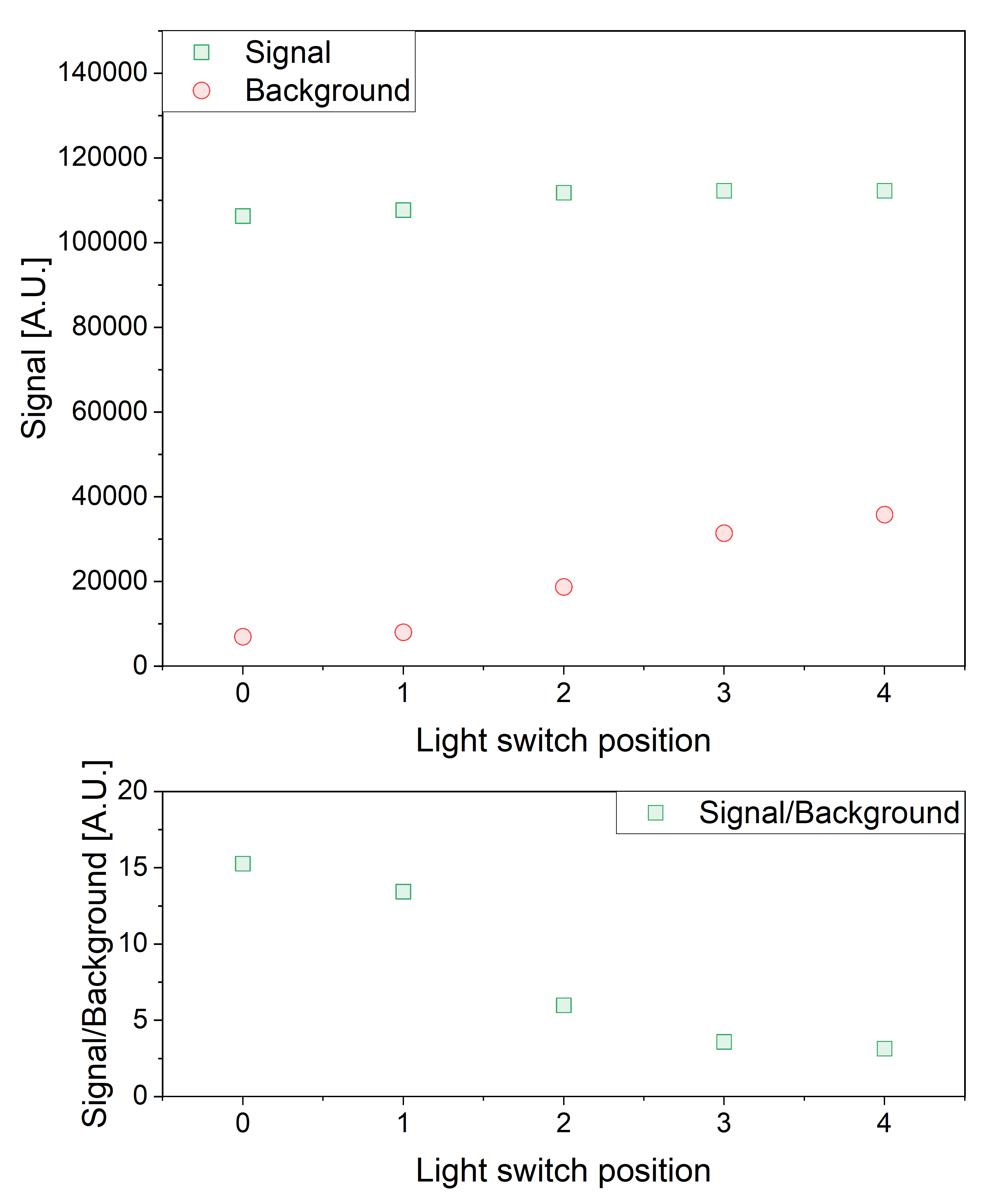}
        \caption{\centering\label{fig: lighton_Cblue}}
    \end{subfigure}
    \begin{subfigure}{.5\textwidth}
        \centering
        \includegraphics[width=\linewidth]{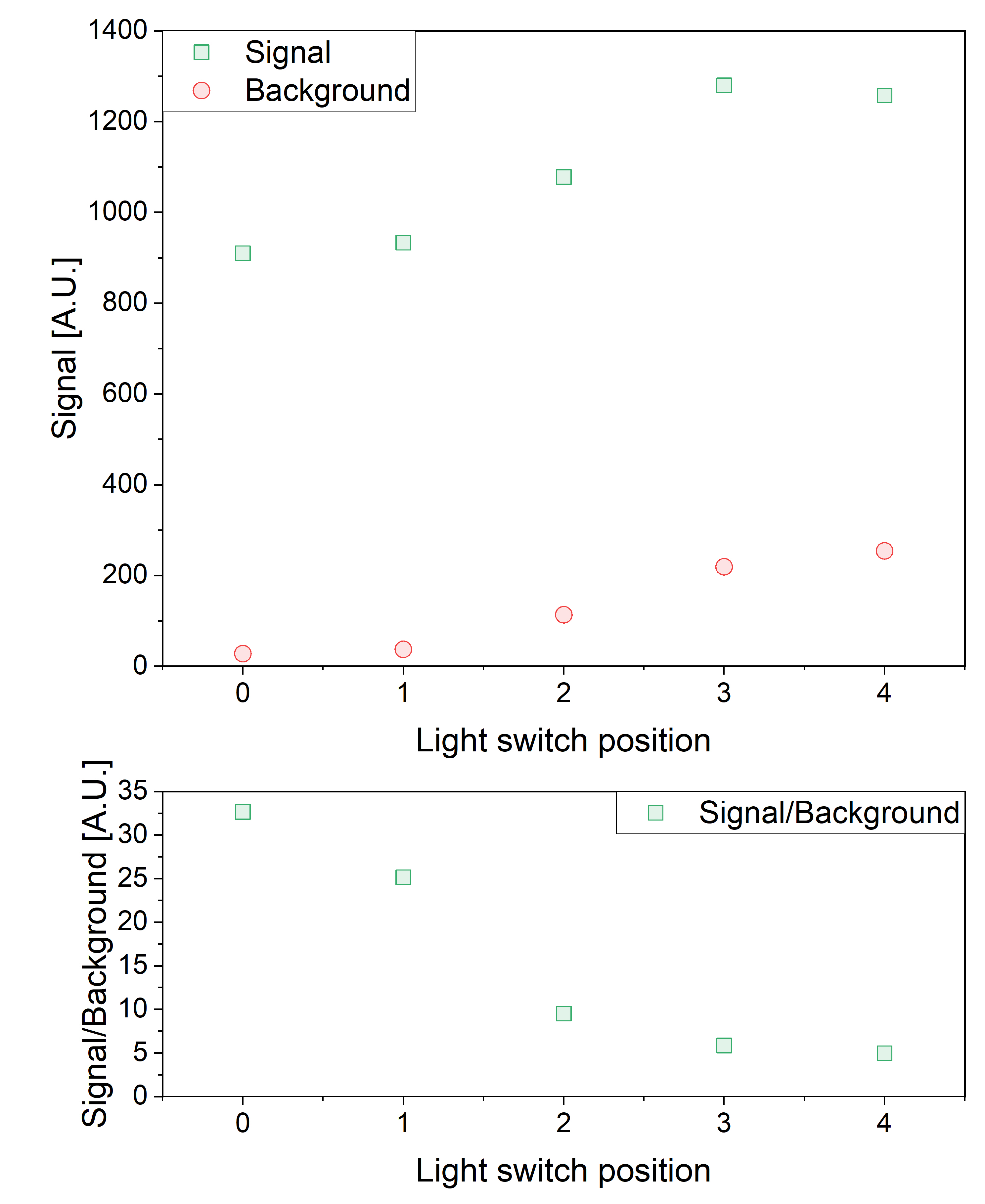}
        \caption{\centering\label{fig: lighton_Basler}}
    \end{subfigure}
    \caption{Signal and background signal variation with ambient light intensity, collected by (a) C-blue and (b) Basler cameras.\label{Fig: lighton}}
\end{figure}

Finally, the scintillating sheet-camera system was tested under conditions of gradually increasing ambient light conditions. It can be seen from Figure~\ref{Fig: lighton} that the intensity of ambient light only affects the background for the C-blue camera, but affects both the signal and the background in case of the Basler camera. It should be noted that the signal of the C-blue saturated for light positions 3 and 4, meaning that a potential signal increase could not be detected. For these 2 light switch positions, the signal to background ratio dropped from 18.7 to 3.1 and from 32.6 to 4.9 for the C-blue and Basler, respectively.

\subsection{Use in a preclinical environment}
\begin{figure}[!ht]
    \centering
    \includegraphics[width=\linewidth]{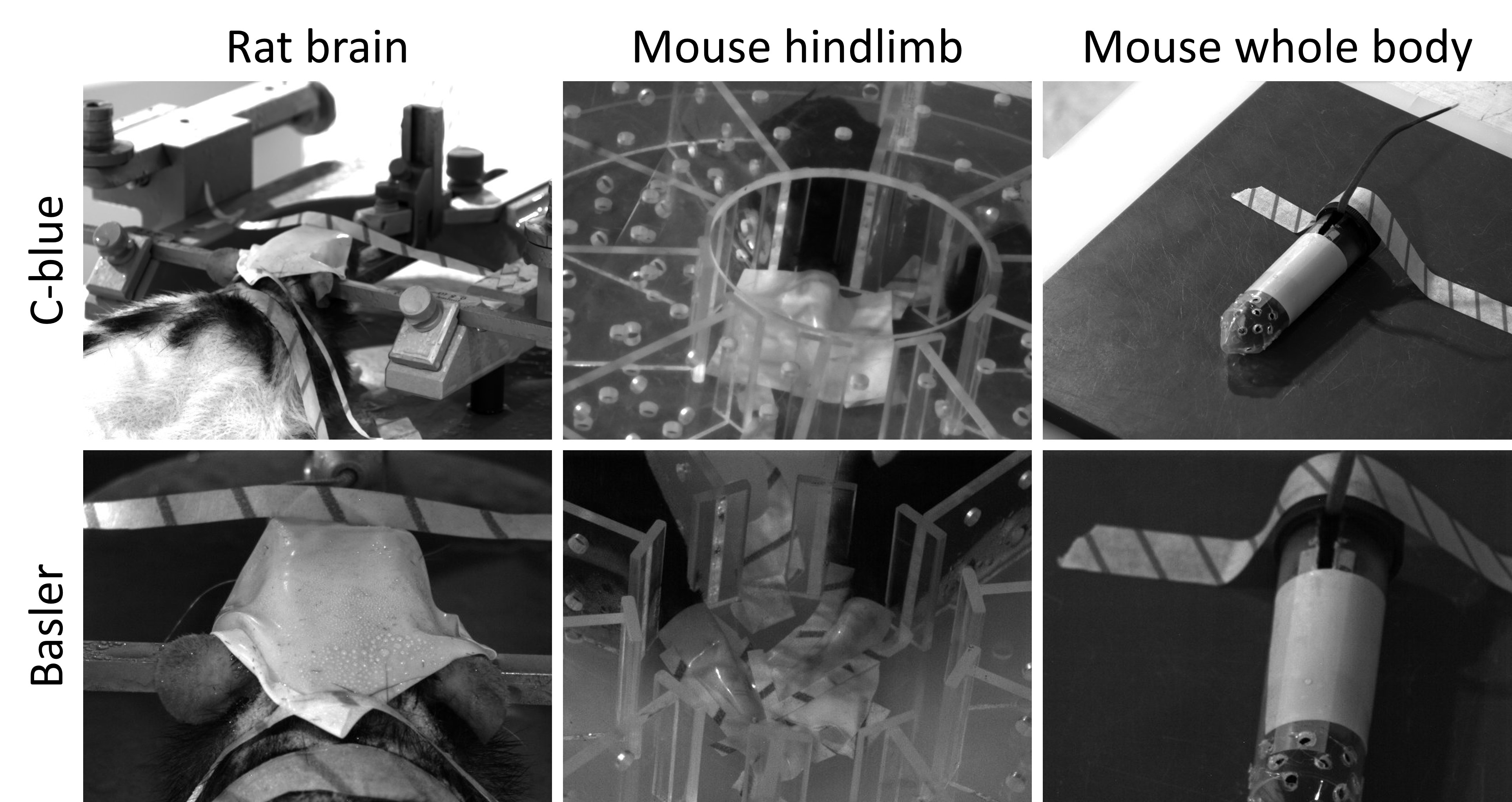}
    \caption{The ex vivo setups to mimic a rat brain, mouse hindlimb and whole body-mouse irradiation from the viewpoint of the C-blue and Basler cameras with the scintillating sheets in position. The rat was positioned using a stereotactic frame. For the hindlimb irradiations, two mice were positioned in a circular holder with their hindlimbs stretched in a central opening. For the whole body mouse irradiations, the mouse was put in a cylindric holder with holes in the front part to allow breathing during in vivo irradiations. \label{Fig: ex vivo position}}
\end{figure}

\begin{figure}[!ht]
    \centering
    \includegraphics[width=\linewidth]{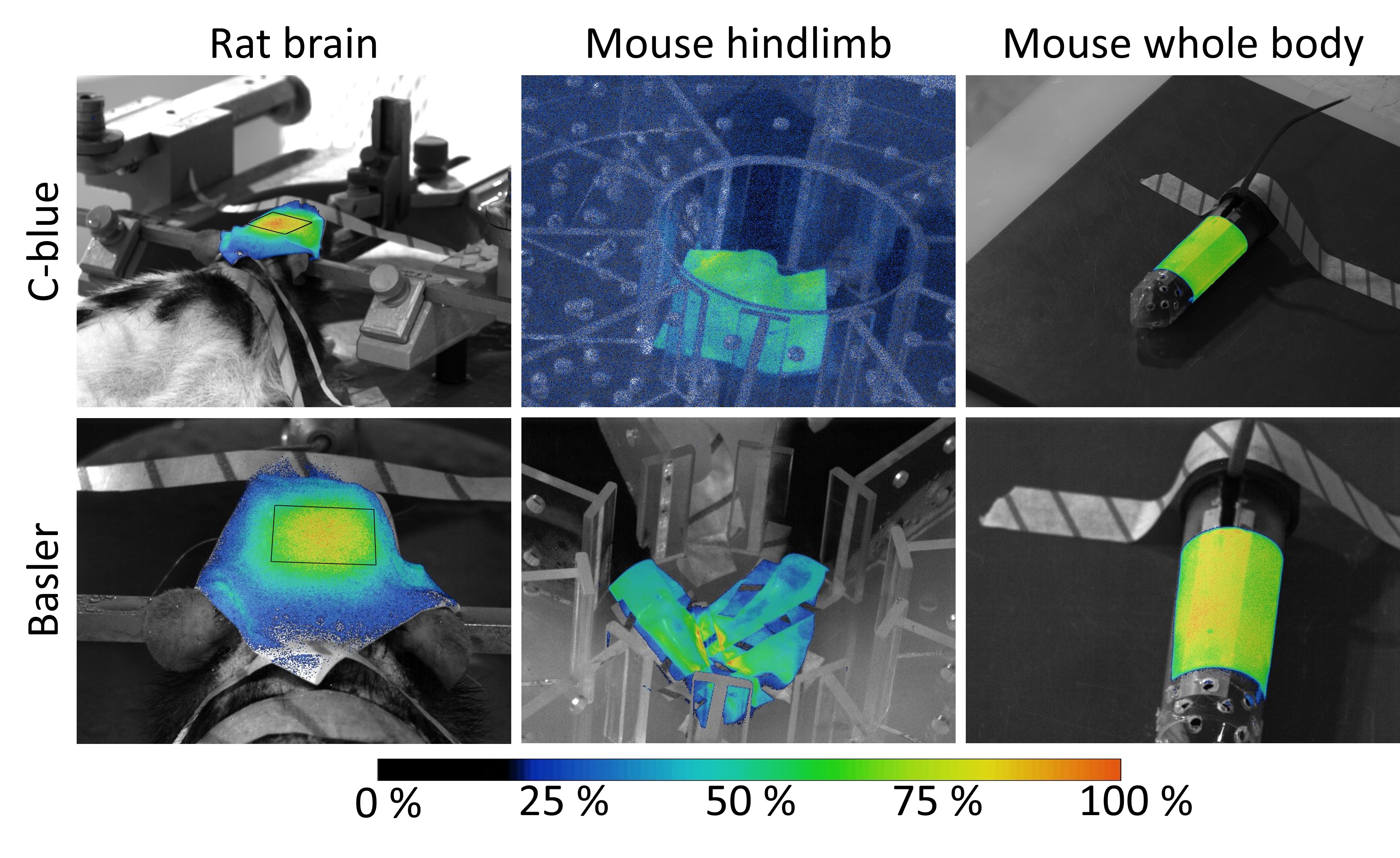}
    \caption{The normalized signal corresponding to a single pulse of an irradiation of a rat brain (DPP = 0.40 Gy), 
 and mouse hindlimb (DPP = 0.93 Gy) using sheet "sample 2" and of a whole mouse brain (DPP = 0.21 Gy) using sheet "sample 1" in combination with the C-blue and Basler cameras.\label{Fig: ex vivo signal}}
\end{figure}

This system was used in 3 different ex vivo situations, representative for preclinical experiments. The different setups are shown in Figure~\ref{Fig: ex vivo position}. The scintillating sheets were molded on the rat skull, mice hindlimbs and whole-body mouse holder and follow the respective curvatures. The signal was superimposed on the positional image as shown in Figure~\ref{Fig: ex vivo signal}. The black square on the rat brain images represents the outline of the field size. It can be seen from all images that the camera position determined the field of view. Parts of the sheet are in a blind spot, disallowing signal determination. The mouse hindlimb results from the C-blue and Basler camera come from two different irradiations. In the first, the entire sheet was used to cover 2 hindlimbs, whereas in the second, the sheet was cut to better match the contour of the individual hindlimbs. Also, the C-blue mouse hindlimb result has a noisy background due to a small iris opening. The "sample 1" sheet is semi transparent, showing a white band from the holder in the whole-body mouse irradiations. The signal from this band is enhanced which is not associated with additional dose delivery, but due to the additional reflectivity of the white band. 
\newline
The linearity with the number of pulses on a curved surface, in a representative preclinical setting was validated using a fixed ROI on the rat's head. The linearity was preserved with an R² of 0.997 and 0.998 for the C-blue and Basler camera, respectively. Despite the difference in setup, the ramp up correction from the characterization was used.
\newline
The introduction of bolus material (supplementary Figure~\ref{Fig: ex vivo Bolus}) complicated the direct overlay of the signal with the positional image. The angled view resulted in a positional shift and cropped the field at the edge of the bolus. In addition, scattering of the light in the bolus material blurred the 2D signal distribution. The light was projected to the flat surface of the bolus. Therefore, the light coming from blind spots was detected. This can be appreciated when comparing the "No bolus" and "1 cm bolus" images of the C-blue camera. The "1 cm bolus" image had a concentric dose distribution, whereas this was deformed in the "No bolus" image. 
\newline
Reduction of field size (supplementary Figure~\ref{Fig: ex vivo Field size}) resulted in a decrease in signal amplitude. With the C-blue camera, the signal was sufficient to discriminate a distinct field, whereas this was not the case for the Basler camera for the 0.6x0.6 and 0.7x0.7 cm² field sizes. The signal of the Basler camera was, in addition, more noisy compared to the one from the C-blue camera. 
\newline
The rat was rotated to assess the robustness of the camera position (supplementary Figure~\ref{Fig: ex vivo Rotation}). The signal was stable within 4.5\% and 3.2\% for the C-blue and Basler camera, respectively. For the C-blue camera, one outlier was excluded as the signal deviated 25.8\% from the average (supplementary Figure~\ref{Fig: Rotation}). 
\newline
The repeatability was assessed by 5 consecutive irradiations of 2 mouse hindlimbs using the same irradiation parameters. The signal was stable within 1.6\% and 1.7\% for the C-blue and Basler camera, respectively, regardless of which leg was investigated.

\newpage

\section{Discussion}
\subsection{Setup optimization}
\label{sec: Discussion setup optimization}
The representative free running and triggered time traces in Figure~\ref{fig: TriggervsNonTrigger} show the benefits and drawbacks of both settings. In free running mode, the relative position of the pulse in the integration window varies which was reflected in an increased variability of the light output of the scintillating sheet. In addition, the non-negligible dead time of the camera, compared to the pulse length, resulted in a significant number of pulses not being detected which is in agreement with the work of Vanreusel et al.~\cite{vanreusel2024dose}. The insert of the figure shows that the decay time of the sheets exceeded the inter-pulse time, which resulted in signal overlap and increased background with pulse number. This is unexpected as the decay time of Ce-doped YAG is in the order of 100 ns~\cite{xia20173+, Moszynski1994}. Further investigation is needed to characterize the decay and afterglow properties of the sheet. A plausible hypothesis is that the elongated signal was due to the interaction with the silicon matrix.
This effect was reflected in a signal increase with pulse number for triggered acquisition.
Here, the relative timing of the integration window to the pulse was the same for every pulse, and the signal rise was characterized and corrected. While significant for "sample 1" and "sample 2", with a signal rise of about 5\%, this signal rise was negligible for "sample 3", being \textless 1\%.
In triggered acquisition the signal was significantly lower than in free running acquisition. This is because the cameras were triggered by the radial readout signal from the monitor chambers and the trigger delay time exceeded the pulse length with a factor \textgreater2. Therefore, light was only collected during the decay of the scintillator. For this reason, the integration window was extended from 1 to 2.5 ms in triggered acquisitions. Another drawback of triggered acquisition was the lack of a time component. In the free running time trace (Figure~\ref{fig: TriggervsNonTrigger}~), it can be seen that the pulses of the Varian Trilogy™ were not equally spaced. This information is lost when acquiring the signal in triggered mode.
\newline
Free running acquisition was beneficial for aDR determination, whereas the triggered acquisition was beneficial for dose, and dose per pulse (DPP) determination. Since triggered acquisition had sufficient signal-to-noise ratio (\textgreater 10) and missing pulses was detrimental for accurate dosimetry, triggered acquisition was utilized for the characterization and tests in the preclinical environment.
\newline
The signal cross over, expressed by a ramp up in the time trace was corrected for as described in section~\ref{sec: preprocessing and analysis}. The introduced correction was shown to remove the ramp up and resulted in a stable signal if no saturation was present in the representative normalized time trace. However, care should be taken to update the correction function when settings such as gain and iris opening are adjusted.

\subsection{Characterization}
This work validates and complements the findings of Vanreusel et al.~\cite{vanreusel2024dose} and Rieker et al.~\cite{Rieker2024}. In Vanreusel et al., a similar scintillating sheet ("sample 3") was characterized, with the C-blue camera in free running mode, in a multicentric study. The dose linearity and independence of aDR, DPP, pulse repetition frequency (PRF) and energy independence were investigated. Vanreusel et al. demonstrated dose linearity up to at least 13 Gy, and independence of aDR, DPP and PRF up to 140 Gy/s, 1.89 Gy and 245 Hz, respectively, and an energy response variation of 3.4\% between 5-9 MeV. In Vanreusel et al., a different calibration was needed for each center, which was attributed to differences in phantom transparency.
In Rieker et al.~\cite{Rieker2024}, a dosimetry method based on a scintillating sheet and integrating current transformer (ICT) was introduced for UHDR VHEE dosimetry. For this approach, the charge density was determined by the total charge ICT and the YAG sheet, and transformed to dose via a calibration factor. Three methods, one analytical and two numerical, were presented to determine the calibration factor. Comparing the dose to 100 regions of interest between their system and radiochromic film, gave root means squared prediction errors of ± 5\%. The accuracy of these methods was found to be sensitive to the camera settings and light output.
These studies showed that this system has the potential to become an adequate 2D real time UHDR dosimeter. However, the use of a converted clinical linacs is very different and the calibration is not trivial. Converted linacs typically have a Gaussian beam profile because the target and flattening filter are removed from the beam path to obtain an UHDR beam~\cite{lempart2019modifying, rahman2021electron}. In this work, the scintillating sheets and radiochromic film were distributed in the field without overlap and received different doses and dose gradients for the same irradiation. The position of the scintillating sheets was fixed between irradiations of the same experiment, but the position of the radiochromic film was subject to small variations (\textless 1 cm), which complicates repeatable dose assessment. Therefore, the number of pulses was used as a surrogate measure for total dose, and radiochromic film dose was used to check for discrepancies in dose delivery. For doses above 22 Gy, the radiochromic film measurements could not be used due to saturation. 
\newline
We hypothesize that the change in signal per pulse for different number of pulses delivered, as observed in Figure~\ref{fig: RampUpAfterCorrection} and~\ref{fig: RampUpAfterCorrection_Basler}, was due to temporal sensitization of the sheet. In this study, the number of pulses were varied ascending, therefore, higher number of pulses resulted in higher signal per pulse. When plotting the signal per pulse of the first pulse for each irradiation against time, a signal increase can be seen when subsequent irradiations are less than 5 minutes apart. When the time between subsequent irradiations is longer, the sensitization effect is reduced (Figure~\ref{Fig: sensitization} in supplementary). Preliminary data (not shown) where the sequence of irradiations was changed supports this hypothesis. Bilski et al. showed that Ce doped YAG crystals also have optically stimulated luminescence (OSL) properties~\cite{Bilski2022}. Therefore, it is plausible that the sensitization is caused by a competition between scintillation and shallow or dosimetric OSL traps, which decreases when these traps get filled~\cite{denis2011influence, nascimento2015short, choudhury2014shallow}. 
\newline
The dose response of the scintillating sheet-camera system was strongly influenced by the setup and camera settings. It can easily be seen that variation of iris opening, focus and integration window changed the amount of photons collected by the CMOS detector. Also, variation of the camera-sheet distance and/or angle, or the presence of light scattering media such as bolus material, resulted in more subtle, non-negligible changes in the amount of collected photons per pixel. In addition, the signal was digitally changed by the gain setting. The choice of these settings and validation of the applicability is paramount. In fact, the observed saturation for the C-blue camera could have been evaded by reduction of the gain or partial closure of the iris opening. Also, the sudden deviation from linearity with the inverted squared SSD of the last 3 data points in Figure~\ref{fig: SSD_Cblue} can be attributed to the change in camera-sheet distance associated with the SSD change. Finally, it was shown that the presence of ambient light affected the performance of the system, without deteriorating it. Especially in conditions when the environmental light was faint, the signal remained an order of magnitude higher than the background, allowing for high fidelity with the use of the system.

\subsection{Use in a preclinical environment}
The system showed potential for surface dosimetry in a preclinical setting, which is an unmet need for UHDR experiments. Preclinical experiments are typically performed using a robust and reproducible setup. The rotating rat experiment was a representation of the setup variability within such an experiment and showed that the dosimetry system was robust within at least 5\%. In this, it is assumed that the dose delivered to each pixel of the sheet was exactly the same for each irradiation. However, this assumption was unlikely to be valid because the Gaussian dose distribution output varies for the linac as does the backscatter. Next to the robustness for setup variability, also stability of the dosimetric system itself is of importance. This was shown by the repeatability experiment at two distinct positions of the sheet. The linearity with pulses experiment showed that the dose-response linearity of the system was preserved in a preclinical setting. In addition, the pulses can be distinguished and a signal per pulse can be obtained. After correcting for the ramp up, this can be used to validate the pulse-to-pulse variability. The use of bolus material, which is often needed in small animal irradiations with high energy electron beams, complicates the projection of the signal back on the patient. It results in a blurred, positionally shifted dose distribution. In addition, the transparency of the bolus may vary during an experimental campaign, which will change the signal distribution. 
\newline
Therefore, in the current confirmation with a single camera and need for case specific calibration, the use of the scintillating sheet-camera system was limited to relative dosimetry, without the use of bolus.

\vspace{2ex}
In previous work from Vanreusel et.al.~\cite{vanreusel2024dose} a similar setup and sheet were characterized and calibrated using free running acquisition in a flat beam. While the calibration showed to be robust within a single setup, the absence of triggering and the correction used to anticipate missed pulses, limited its application in preclinical and clinical settings. This work showed the benefits of triggering, limiting the signal variability due to the temporal position of the integration window with respect to the pulse. In addition, the present study illustrated the need and challenges for 2D dosimetry in non-homogeneous beams. The system was shown to be useful, provided that it was recalibrated when changing the setup and/or camera settings and that the time between subsequent irradiations was sufficient such that sensitization becomes irrelevant. A single calibration irradiation per setup should be sufficient. It was also shown that this system can be used in the presence of ambient light for UHDR dosimetry, although it should be minimized for optimal performance. In the clinics, the room light is typically dimmed for laser-based patient positioning. This setting could serve as compromise between patient comfort and dosimetric performance of this system.

\subsection{Future perspectives}
While the system was shown to demonstrate potential, there remain improvements prior to its routine application for absolute dosimetery in preclinical (UHDR) experiments. The decay time of the material was a blessing and a curse. Sufficient signal, 10 µs post pulse, was needed to allow triggered data acquisition, but the decay should not exceed the inter-pulse time since this lead to a ramp up of the signal. Alternative scintillating materials should be investigated to optimize this trade-off. Simultaneously, an alternative trigger source should be taken such that the integration window coincides with the pulse, allowing for the use of materials with very short (ns) decay times. 
\newline
An additional time component should be introduced for triggered acquisition to allow post irradiation aDR assessment. Machine log files or image timestamps with ms time resolution could meet this requirement.
\newline
The configuration with a single camera was challenging for the complex surfaces encountered in radiotherapy treatments. Also, the use of a checkerboard or graph paper to transform the dose distribution to a beam's eye view was insufficient. Therefore, a more robust configuration, including multiple cameras positioned at different angles would facilitate the need to compensate for blind spots and to allow signal reconstruction from a beam's eye view. Such technology, using 3 cameras (or pods) and patching the information to generate a large 3D field of view, is already frequently used in surface scanning for the positioning of patients. As surface guidance is becoming standard in radiotherapy departments, an integration in this system can be considered. Preferably, also a digital twin of the setup would be constructed such that the dose of the system can be superimposed on it. Rather than a beam's eye view, such a solution allows a 3D rendering. In combination with a treatment planning system (TPS), this would allow patient specific quality control~\cite{Clark2024}.
\newline
In fact, a TPS is needed for the calibration of this 2D system. Due to the flexibility of the sheet, pixel wise cross calibration against other 2D dosimeters is not an option. 
\newline
For further research towards a viable flexible 2D UHDR dosimeter in a preclinical setting, the authors suggest the use of 3D printed animal phantoms rather than ex vivo set ups. This has the benefit of reusability of the same subject, and not being influenced by the conditions of the animal such as hair contamination resulting from shaving prior to study or water contamination due to defrosting from frozen subjects during the course of a study. An initial feasibility test was performed with a 3D printed mouse as shown in supplementary figure~\ref{Fig: 3D mouse}.

\section{Conclusion}
This work investigated the use of a flexible YAG-based scintillating sheet-camera combination for preclinical UHDR research. It showed that triggered image acquisition was required due to the substantial dead time of the camera compared to the pulse length. The decay time of the sheet was shown to be too long, requiring a ramp up correction of the signal in triggered mode. Hereafter, the system was shown to provide an output linear with radiation dose and aDR independent. This dose linearity was preserved in a preclinical setting and after the introduction of bolus material, however recalibration was needed. Its use in preclinical setting is critically needed during UHDR studies since such a system has the potential to yield a real time output of dose. The results presented demonstrate that the system feasible with intrasubject stability \textless5\% and intersubject stability \textless2\%. More optimization is required, especially in processing. Also, alternative materials with shorter decay time need to be investigated, in combination with an optimized triggering signal.

\section*{CRediT authorship contribution statement}
\textbf{Verdi Vanreusel}: Conceptualization, Methodology, Software, Validation, Formal analysis, Investigation, Data Curation, Writing - Original Draft, Visualization;
\textbf{Steve Brown}: Conceptualization, Validation, Investigation, Resources, Data Curation, Writing - Review \& Editing, Supervision, Project administration, Funding acquisition;
\textbf{Shujat Ali}: Validation, Investigation, Data Curation, Writing - Review \& Editing;
\textbf{Thomas De Kerf}: Methodology, Software, Resources; 
\textbf{Anthony J. Doemer}: Resources;
\textbf{Paul Leblans}: Resources; 
\textbf{Benjamin Movsas}: Resources, Funding acquisition; 
\textbf{Humza Nusrat}: Resources;
\textbf{Behzad Shirmard}: Resources; 
\textbf{Kundan Thind}: Resources, Project administration; 
\textbf{Steve Vanlanduit}: Methodology, Software, Resources;
\textbf{Verellen Dirk}: Resources, Writing - Review \& Editing, Supervision, Project administration, Funding acquisition;
\textbf{Gasparini Alessia}: Writing - Review \& Editing, Supervision; and 
\textbf{Luana de Freitas Nascimento}: Writing – review \& editing, Validation, Supervision, Project administration, Methodology, Investigation, Data curation, Conceptualization.

\newpage
\printbibliography
\newpage
\appendix
\setcounter{table}{0}
\setcounter{figure}{0}
\renewcommand{\thefigure}{A\arabic{figure}}

\section{Supplementary}

\begin{figure}[!ht]
    \centering
    \includegraphics[width=\linewidth]{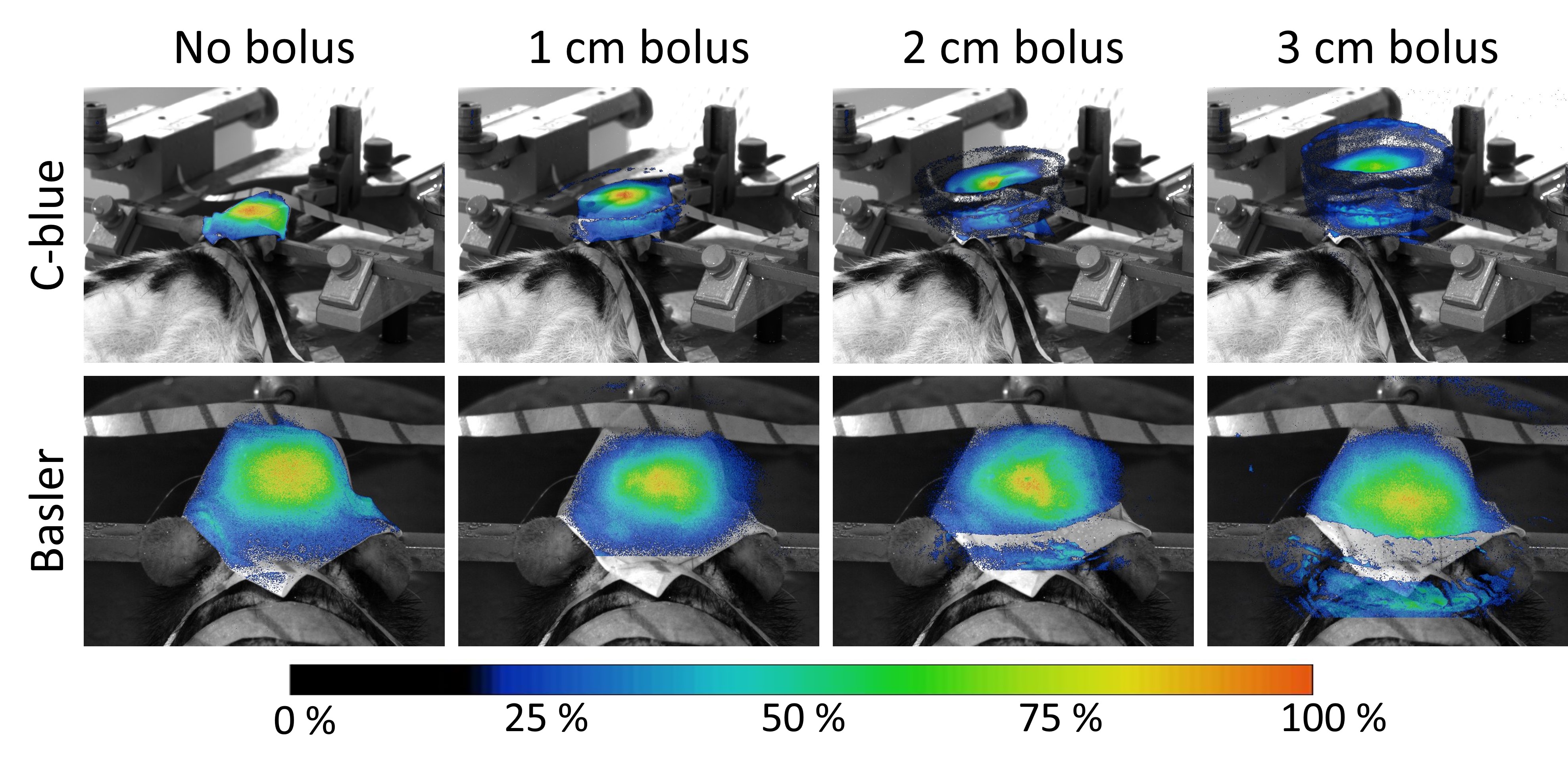}
    \caption{The recording of a rat brain irradiation with different bolus' thickness.\label{Fig: ex vivo Bolus}}
\end{figure}

\begin{figure}[!ht]
    \centering
    \includegraphics[width=\linewidth]{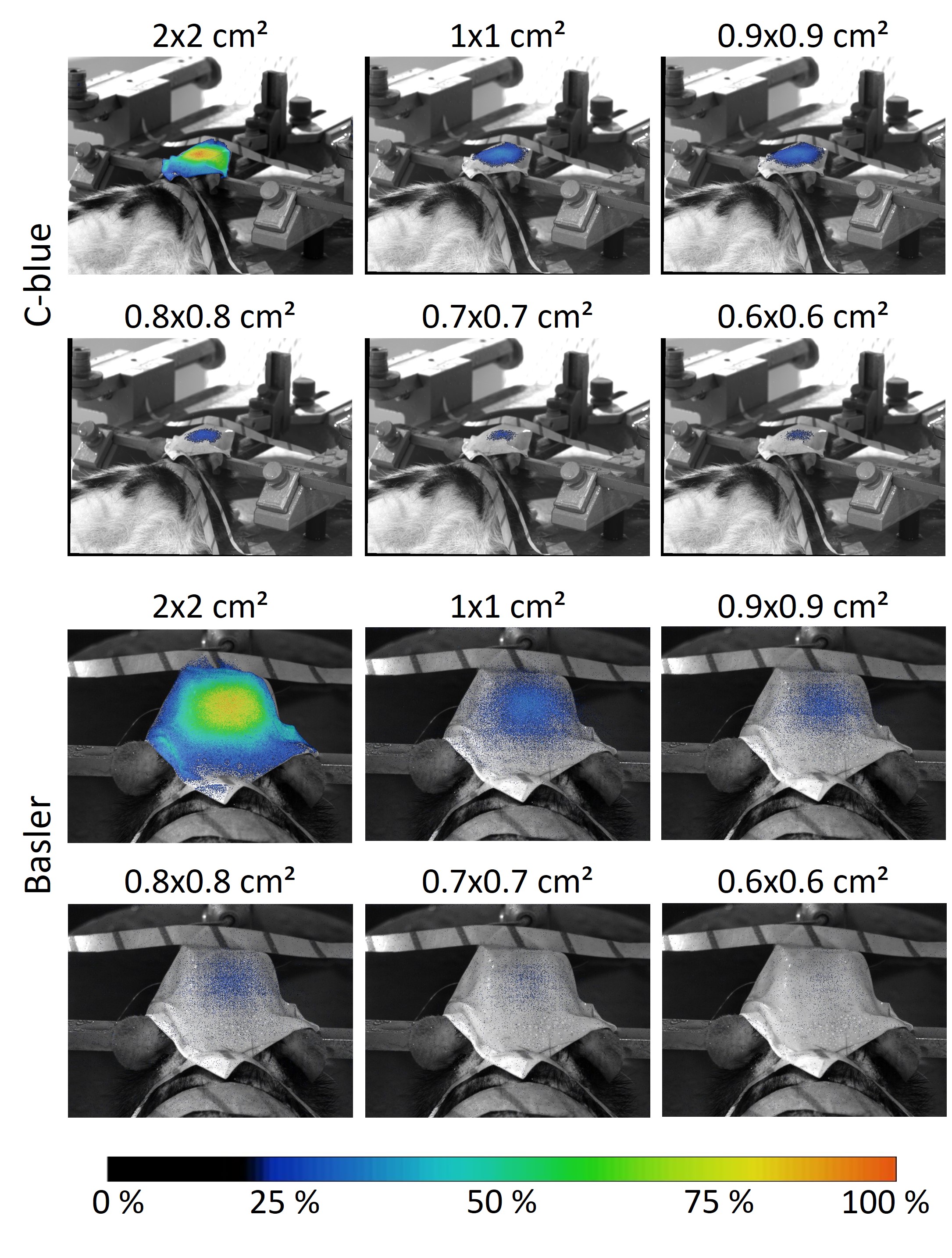}
    \caption{The recording of a rat brain irradiation for different field sizes.\label{Fig: ex vivo Field size}}
\end{figure}

\begin{figure}[!ht]
    \centering
    \includegraphics[width=\linewidth]{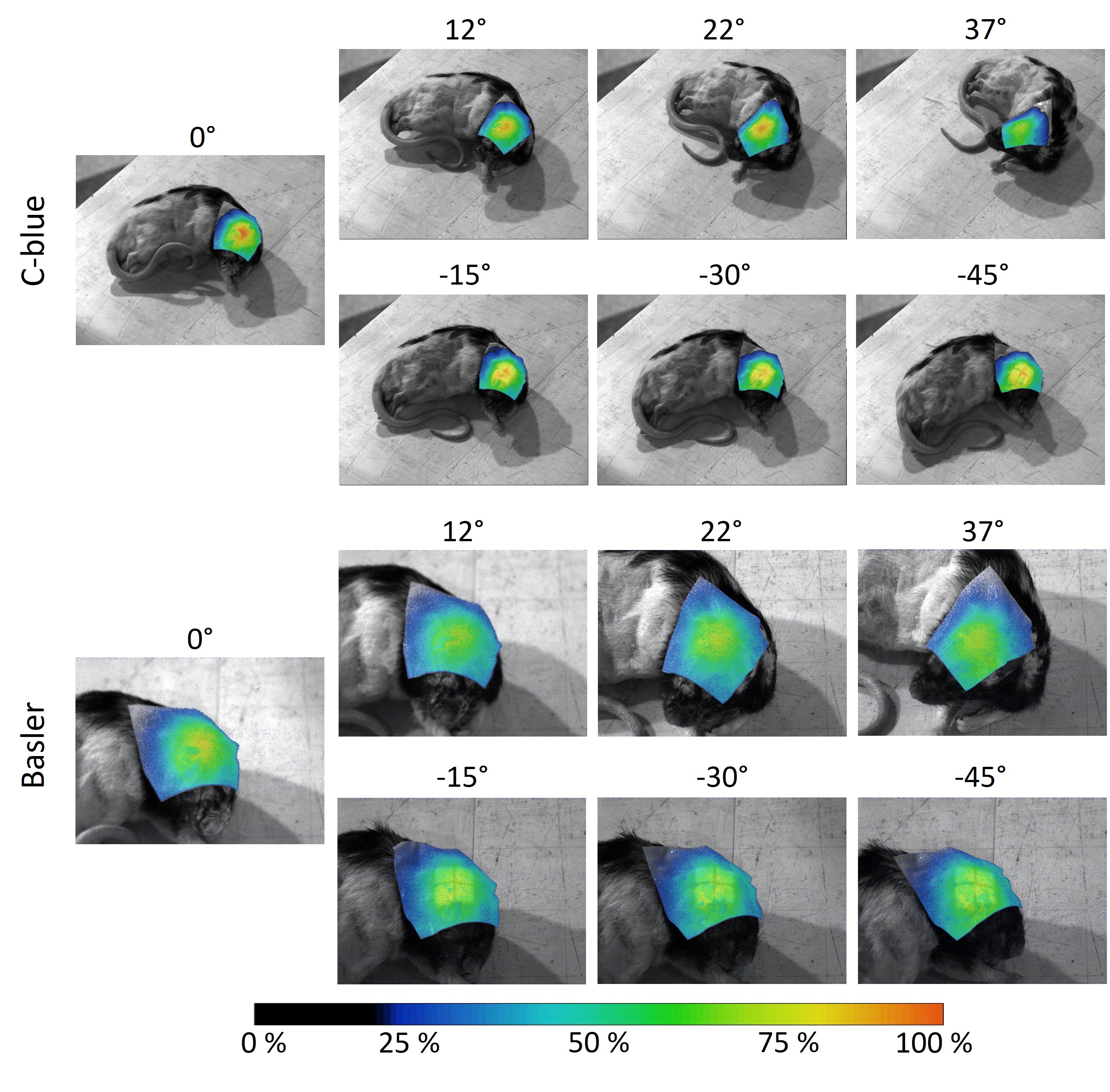}
    \caption{The recording of a rat brain irradiation for different rotation angles.\label{Fig: ex vivo Rotation}}
\end{figure}

\begin{figure}[!ht]
    \begin{subfigure}{.5\textwidth}
        \centering
        \includegraphics[width=\linewidth]{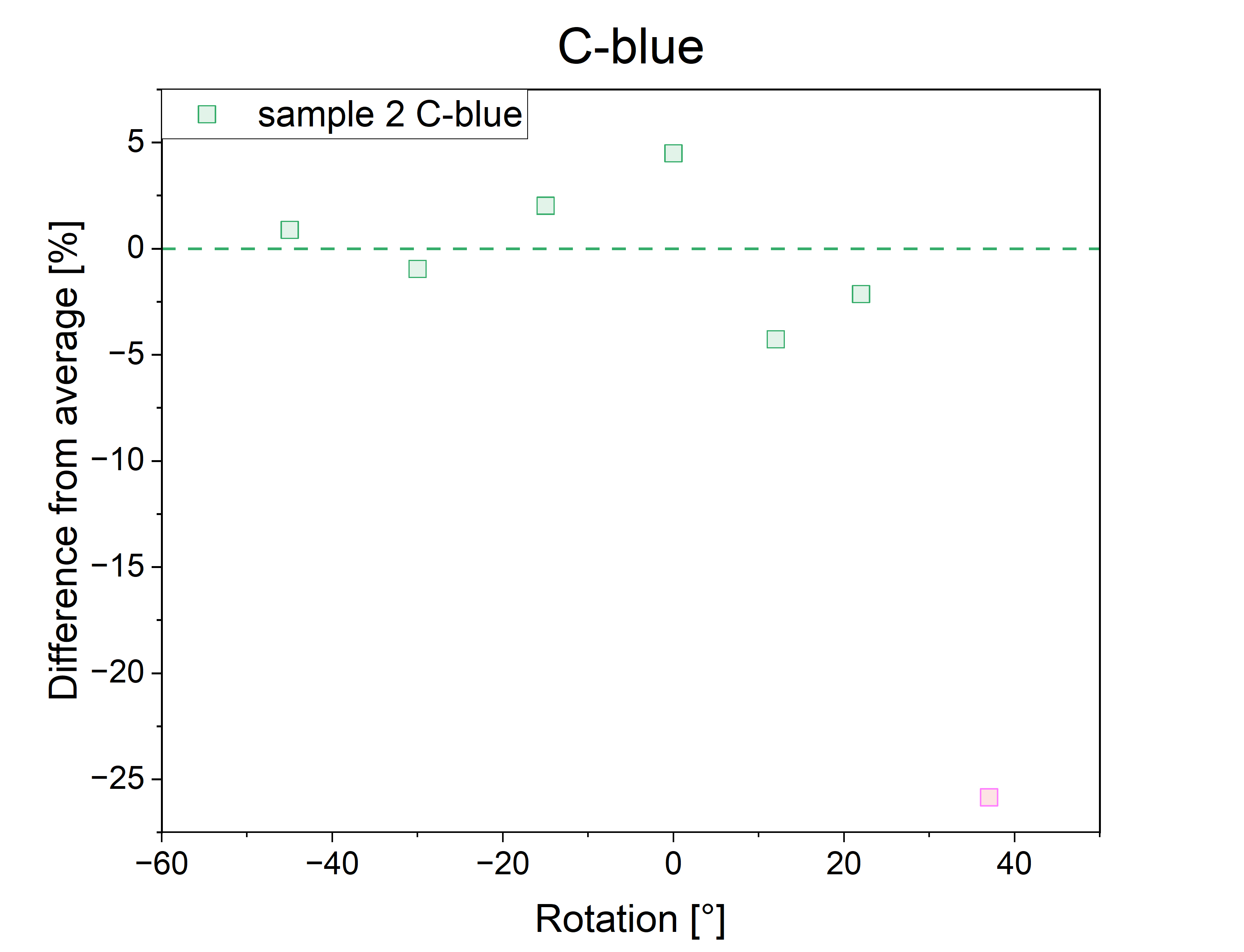}
        \caption{\centering}
    \end{subfigure}
    \begin{subfigure}{.5\textwidth}
        \centering
        \includegraphics[width=\linewidth]{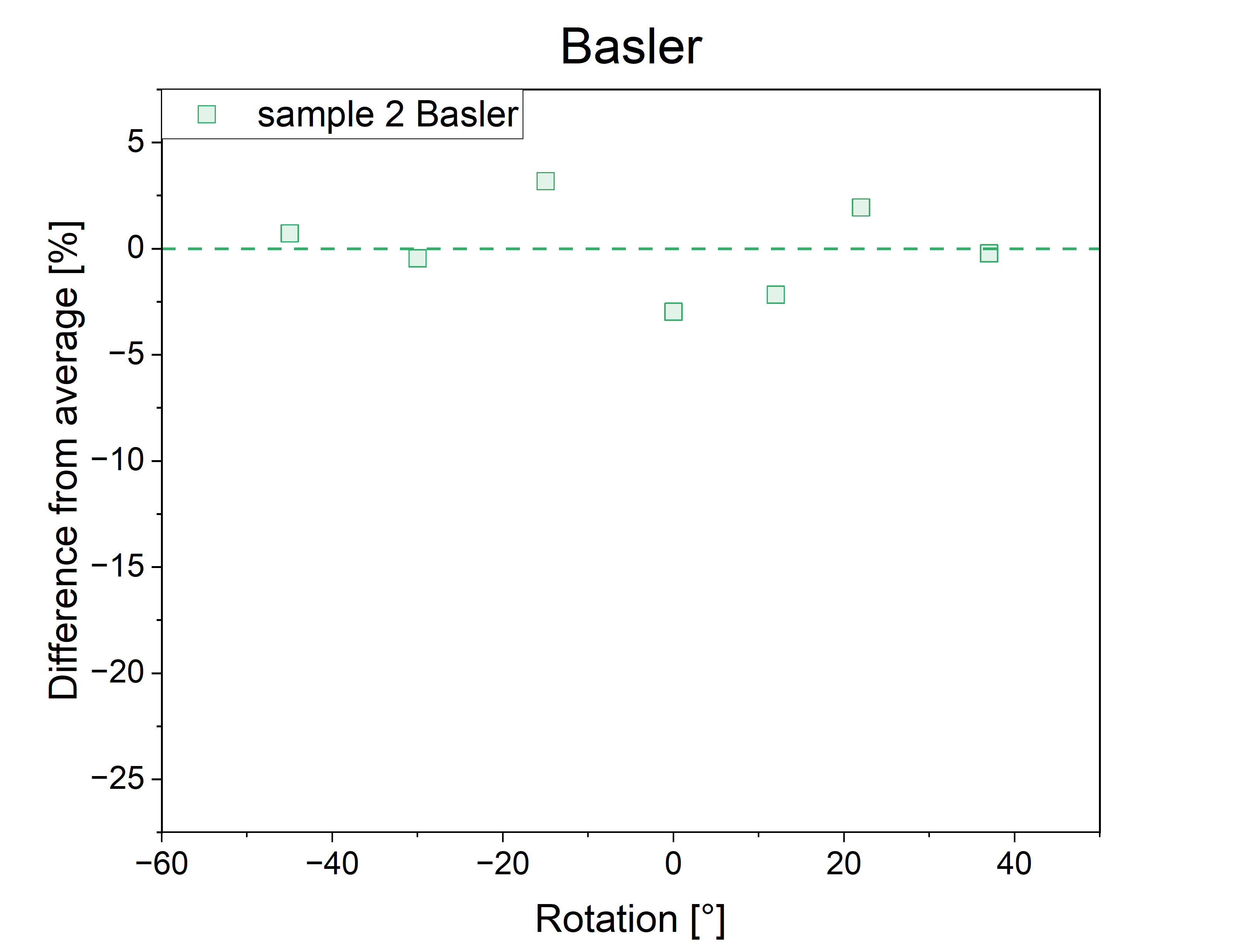}
        \caption{\centering}
    \end{subfigure}
    \caption{The signal stability as function of the rotation of the ex vivo rat to assess the robustness of the setup.\label{Fig: Rotation}}
\end{figure}

\begin{figure}[!ht]
    \centering
    \includegraphics[width=\linewidth]{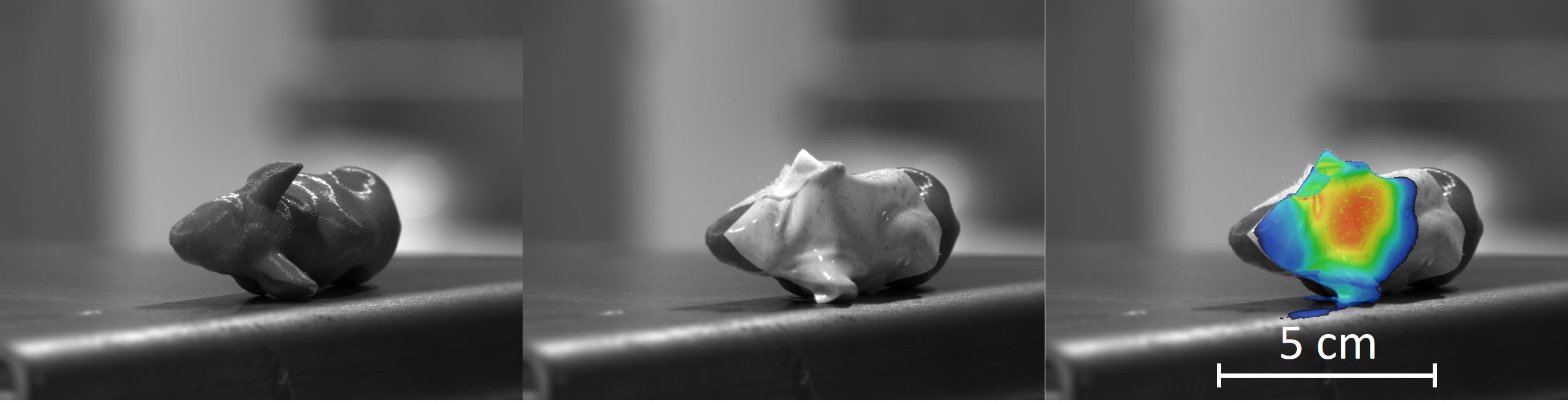}
    \caption{The feasibility test of a setup using a 3D printed mouse.\label{Fig: 3D mouse}}
\end{figure}

\begin{figure}[!ht]
    \centering
    \includegraphics[width=\linewidth]{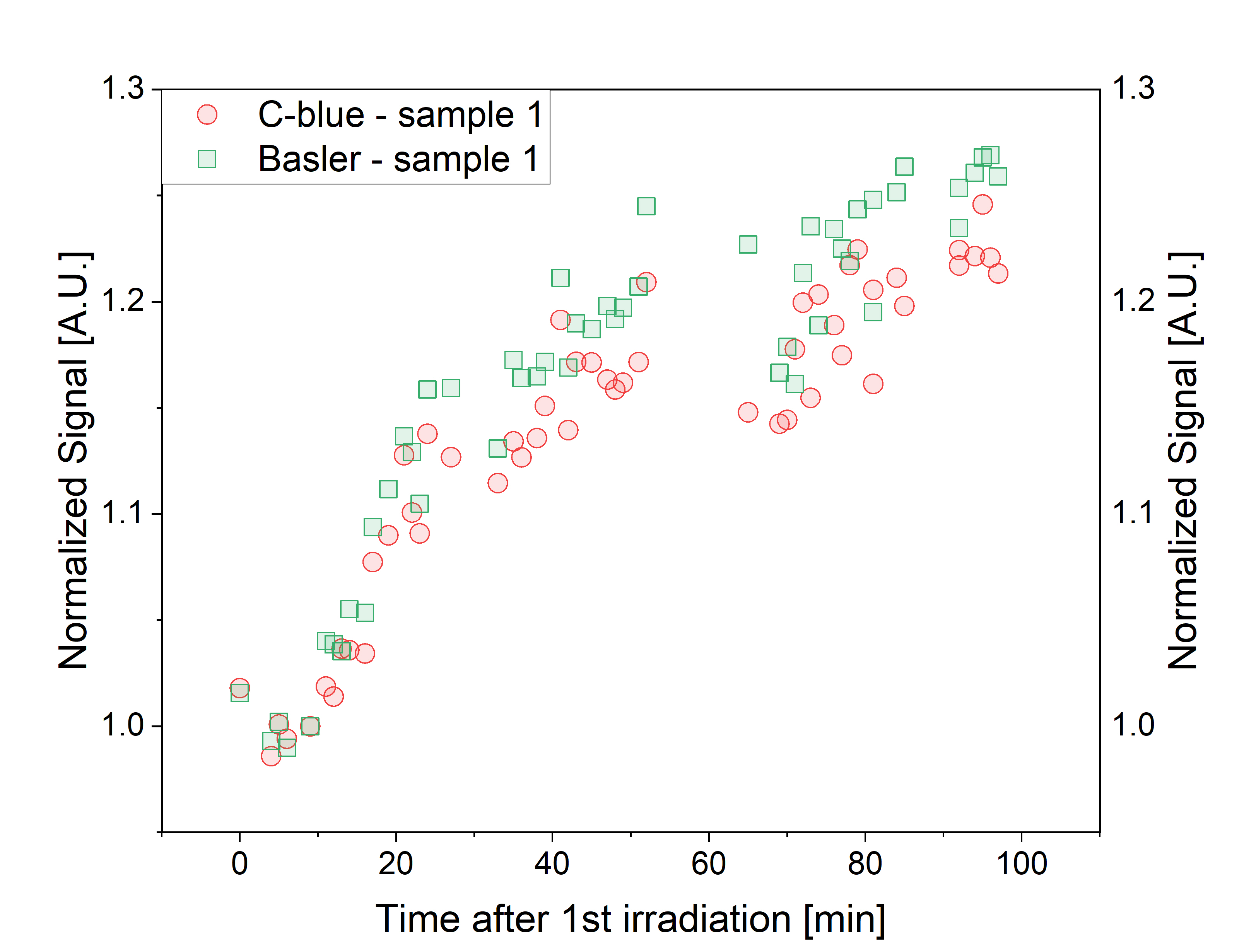}
    \caption{The signal per pulse of the first pulses of the irradiations from the linearity with dose experiment against time, showing the sensitization of the sheet.\label{Fig: sensitization}}
\end{figure}

\end{document}